\documentclass[useAMS,usenatbib]{mn2e}
\usepackage{graphicx}
\usepackage{times}
\usepackage{xfrac}
\usepackage[T1]{fontenc}
\usepackage{aecompl}

\newcommand{\Ion}[2]{#1{\,\sc#2}}
\newcommand{\kms}{\mbox{$\mathrm{km\,s^{-1}}$}}
\newcommand{\MSUN}{\mbox{$\mathrm{M_{\odot}}$}}
\newcommand{\RSUN}{\mbox{$\mathrm{R_{\odot}}$}}

\title[The eclipsing WD/dM binary QS Vir]{The crowded magnetosphere of the post common envelope binary QS\,Virginis}

\author[S. G. Parsons et al.]{S.~G.~Parsons$^{1}$\thanks{steven.parsons@uv.cl},
C.~A.~Hill$^{2,3}$,
T.~R.~Marsh$^{4}$,
B.~T.~G{\"a}nsicke$^{4}$,
C.~A.~Watson$^{2}$,
\newauthor
D.~Steeghs$^{4}$,
V.~S.~Dhillon$^{5,6}$,
S.~P.~Littlefair$^{5}$,
C.~M.~Copperwheat$^{7}$,
\newauthor
M.~R.~Schreiber$^{1,8}$
and M.~Zorotovic$^{1}$
\\
$^{1}$ Departmento de F{\'i}sica y Astronom{\'i}a, Universidad de
Valpara{\'i}so, Avenida Gran Bretana 1111, Valpara{\'i}so, 2360102, Chile\\
$^{2}$ Astrophysics Research Centre, Queen's University Belfast, Belfast, BT7
1NN Northern Ireland, UK\\
$^{3}$ IRAP, Observatoire Midi-Pyr{\'e}n{\'e}es, University of Toulouse, 14
avenue Edouard Belin, 31400, Toulouse, France\\
$^{4}$ Department of Physics, University of Warwick, Coventry CV4 7AL, UK\\
$^{5}$ Department of Physics and Astronomy, University of Sheffield, Sheffield, S3 7RH, UK\\
$^{6}$ Instituto de Astrof{\'i}sica de Canarias, V{\'i}a Lactea s/n, La Laguna, E-38205 Tenerife, Spain\\
$^{7}$ Astrophysics Research Institute, Liverpool John Moores University, IC2, Liverpool Science Park, 146 Brownlow Hill, Liverpool L3 5RF\\
$^{8}$ Millenium Nucleus "Protoplanetary Disks in ALMA Early Science", Universidad de Valparaiso, Valparaiso 2360102, Chile}
\begin{document}
\input{references.cls}
\date{Accepted 2016 March 1. Received 2016 March 1; in original form 2015 November 16}

\pagerange{\pageref{firstpage}--\pageref{lastpage}} \pubyear{2015}

\maketitle

\label{firstpage}

\begin{abstract}

We present high speed photometry and high resolution spectroscopy of the
eclipsing post common envelope binary QS\,Virginis (QS\,Vir). Our UVES spectra
span multiple orbits over more than a year and reveal the presence of several
large prominences passing in front of both the M star and its white dwarf
companion, allowing us to triangulate their positions. Despite showing small
variations on a timescale of days, they persist for more than a year and may
last decades. One large prominence extends almost three stellar radii from the
M star. Roche tomography reveals that the M star is heavily spotted and that
these spots are long-lived and in relatively fixed locations, preferentially
found on the hemisphere facing the white dwarf. We also determine precise
binary and physical parameters for the system. We find that the $14,220 \pm
350$\,K white dwarf is relatively massive, $0.782 \pm 0.013$\MSUN, and has a
radius of $0.01068 \pm 0.00007$\RSUN, consistent with evolutionary models. The
tidally distorted M star has a mass of $0.382 \pm 0.006${\MSUN} and a radius of
$0.381 \pm 0.003$\RSUN, also consistent with evolutionary models. We find that
the magnesium absorption line from the white dwarf is broader than
expected. This could be due to rotation (implying a spin period of only
$\sim$700 seconds), or due to a weak ($\sim$100kG) magnetic field, we favour
the latter interpretation. Since the M star's radius is still within its Roche 
lobe and there is no evidence that its over-inflated we conclude that
QS\,Vir is most likely a pre-cataclysmic binary just about to become
semi-detached.

\end{abstract}


\section{Introduction}

Close binaries containing a white dwarf and a low-mass M star are survivors
of a common envelope phase of evolution during which both stars orbited
within a single envelope of material ejected by the progenitor of the white
dwarf. These systems slowly lose angular momentum via gravitational radiation
and (if the M star is massive enough) magnetic braking, eventually becoming
semi-detached cataclysmic variable (CV) systems.

\begin{table*}
 \centering
  \caption{Journal of observations. The eclipse of the white dwarf occurs at
    phase 1, 2 etc.}
  \label{tab:obslog}
  \begin{tabular}{@{}lcccccccc@{}}
  \hline
  Date at     &Instrument&Telescope &Filter(s) &Start     &Orbital     &Exposure &Number of &Conditions              \\
  start of run&          &          &          &(UT)      &phase       &time (s) &exposures &(Transparency, seeing)  \\
  \hline
  2002/05/20  & ULTRACAM & WHT      & $u'g'r'$ & 20:51 & 0.48--1.55 & 1.1     & 10533    & Fair, $\sim$2 arcsec   \\
  2003/05/20  & ULTRACAM & WHT      & $u'g'i'$ & 23:43 & 0.93--1.64 & 2.9     & 1983     & Variable, 1.2-3 arcsec \\
  2003/05/22  & ULTRACAM & WHT      & $u'g'i'$ & 00:39 & 0.84--1.10 & 2.9     & 957      & Good, $\sim$1.5 arcsec \\
  2003/05/24  & ULTRACAM & WHT      & $u'g'i'$ & 22:02 & 0.38--1.07 & 2.9     & 2233     & Good, $\sim$1.2 arcsec \\
  2006/03/13  & ULTRACAM & WHT      & $u'g'r'$ & 00:42 & 0.88--1.09 & 2.4     & 1098     & Fair, $\sim$2 arcsec   \\
  2010/04/21  & ULTRACAM & NTT      & $u'g'i'$ & 06:35 & 0.88--1.37 & 1.7     & 3665     & Good, $\sim$1.2 arcsec \\
  2011/01/07  & ULTRACAM & NTT      & $u'g'i'$ & 08:08 & 0.89--1.09 & 1.9     & 1409     & Good, $\sim$1.3 arcsec \\
  2011/05/29  & ULTRACAM & NTT      & $u'g'r'$ & 04:49 & 0.92--1.08 & 3.9     & 550      & Poor, $\sim$1.5 arcsec \\
  2013/05/05  & UVES     & VLT      & -        & 00:10 & 0.28--2.04 & 180.0   & 98       & Excellent, $<$1 arcsec \\
  2013/05/05  & UVES     & VLT      & -        & 23:52 & 0.83--2.15 & 180.0   & 74       & Excellent, $<$1 arcsec \\
  2014/04/24  & UVES     & VLT      & -        & 03:01 & 0.21--1.30 & 180.0   & 62       & Good, $\sim$1.0 arcsec \\
  2014/04/25  & UVES     & VLT      & -        & 02:19 & 0.65--1.74 & 180.0   & 62       & Good, $\sim$1.0 arcsec \\
  2014/05/01  & UVES     & VLT      & -        & 02:04 & 0.38--1.51 & 180.0   & 62       & Good, $\sim$1.3 arcsec \\
  2014/05/31  & UVES     & VLT      & -        & 23:30 & 0.29--1.39 & 180.0   & 62       & Excellent, $<$1 arcsec \\
  \hline
\end{tabular}
\end{table*}

In recent years the number of such systems known has dramatically increased,
thanks mainly to the Sloan Digital Sky Survey (SDSS,
\citealt{york00,adelman08,abazajian09}), which has lead to the discovery of
more than 2000 white dwarf plus main-sequence systems \citep{rebassa13}, from
which more than 200 close, post common-envelope binaries (PCEBs) have been
identified \citep{nebot11,parsons13css,parsons15}. Among this sample are 71
eclipsing binaries (see the appendix of \citealt{parsons15} for a recent
census). These are extremely useful systems for high precision stellar
parameter studies, since
the small size of the white dwarf leads to very sharp eclipse profiles,
allowing radius measurements to a precision of better than two per cent
\citep[e.g.][]{parsons10nn}, good enough to test models of stellar and binary
evolution.

The main-sequence star components in PCEBs are tidally locked to the white
dwarf and hence are rapidly rotating. This rapid rotation results in very
active stars. Indeed, it appears as though virtually every PCEB
hosts an active main-sequence star, regardless of its spectral type
\citep{rebassa13act}. Even extremely old systems still show signs of
activity \citep{parsons12cool}, manifesting as starspots and flaring. These
features allow us to constrain the configuration of the underlying magnetic
field of the main-sequence star, crucial for understanding the evolution of
these binaries, since the magnetic field is able to remove angular momentum
from the system, driving the two stars closer together
\citep{rappaport83,kawaler88}.

Discovered as an eclipsing PCEB by \citet{odonoghue03}, QS\,Vir
(EC\,13471$-$1258) has been intensely studied since it shows signs of being
both a standard pre-CV, detached white dwarf plus main-sequence binary
\citep{ribeiro10,parsons11}, as well as 
indications of accretion of material on to the white dwarf well above standard
rates for detached systems \citep{matranga12}; the M star is also very close
to filling its Roche lobe, leading to the initial interpretation that it could
in fact be a hibernating CV that has recently detached due to a nova eruption
\citep{odonoghue03}. Furthermore, the arrival times 
of its eclipse show substantial variations \citep{parsons10time}, which some
authors have claimed may be due to the gravitational effects of circumbinary
planets \citep{almeida11}, although the proposed orbital configuration has
recently been shown to be highly unstable \citep{horner13}, meaning that the
true origin of these variations remains unexplained. 

The M star in QS\,Vir is very active, indicated both by evidence of
substantial flaring \citep{parsons10time}, as well as intriguing narrow
absorption features detected in high resolution spectra by
\citet{parsons11}. This absorption was found to be due to material expelled
from the M star passing in front of the white dwarf. In this paper we
investigate the behaviour of the material lost by the M star and its motion
within the binary over a time span of more than a year as well as performing
Roche tomography to identify and track starspots on the surface of the M
star. We also place stringent constraints on the stellar and binary
parameters. 

\section{Observations and their reduction}

In this section we outline our observations and their reduction. A full log of
all our observations is given in Table~\ref{tab:obslog}.

\subsection{ULTRACAM photometry}

QS\,Vir has been observed many times with the high speed frame-transfer camera
ULTRACAM \citep{dhillon07}. These observations date back to 2002 and were
obtained with ULTRACAM mounted as a visitor instrument on the 4.2-m William
Herschel Telescope (WHT) on La Palma and the 3.5-m New Technology Telescope
(NTT) on La Silla. Much of these data were presented in \citet{parsons10time},
however, we detail both the older data and our new data in
Table~\ref{tab:obslog}. ULTRACAM uses a triple beam setup allowing one to
obtain data in the $u'$, $g'$ and either $r'$ or $i'$ band simultaneously.

All of these data were reduced using the ULTRACAM pipeline
software. Debiassing, flatfielding and sky background subtraction were
performed in the standard way. The source flux was determined with aperture
photometry using a variable aperture, whereby the radius of the aperture is
scaled according to the full width at half maximum (FWHM). Variations in
observing conditions were accounted for by determining the flux relative to a
comparison star in the field of view. The data were flux calibrated using
observations of standard stars observed during twilight.

\subsection{UVES spectroscopy}

We observed QS\,Vir with the high resolution Echelle spectrograph UVES
\citep{dekker00} installed at the European Southern Observatory Very Large
Telescope (ESO VLT) 8.2-m telescope unit on Cerro Paranal. We used the
dichroic 2 setup allowing us to cover 3760{\AA}--4990{\AA} in the blue arm and
5690{\AA}--9460{\AA} in the red arm, with a small gap between 7520{\AA} and
7660{\AA} and 2x2 binning in both arms. We used exposure times of 3 minutes in
order to reduce the effects of orbital smearing, this gives a signal-to-noise
ratio in the continuum ranging from $\sim$15 at the shortest wavelengths to
$\sim40$ at longer wavelengths. 

Our data consist of two separate observing runs. The first, in 2013, was conducted in visitor mode over two nights. On each night we covered well over a full orbital period. During this run we also observed several M star spectral type standards covering the range M3.0 to M4.5. The second run comprised of a series of four service mode observations made over a month roughly a year after our visitor mode observations. Each of these observations covered just over an orbital period.

All the data were reduced using the most recent release of the UVES data reduction pipeline (version 5.3.0). The standard recipes were used to optimally extract each spectrum. We used observations of the featureless DC white dwarf LHS\,2333 to flux calibrate and telluric correct our spectra.

\section{Results}

\subsection{Spectrum}

The spectrum of QS\,Vir was first described by
\citet{odonoghue03}. \citet{ribeiro10} and \citet{parsons11} also presented
higher resolution spectra. Both components of the binary are clearly visible
in the spectrum, with the Balmer lines of the white dwarf the dominant
features at wavelengths less than $\sim$5000{\AA} and the molecular
absorption features of the M star the dominant features at longer
wavelengths. There are also several emission lines originating from the M star
throughout the spectrum caused by a combination of activity and irradiation
from the white dwarf. As noted by \citet{parsons11} there is also a weak
\Ion{Mg}{ii} 4481{\AA} absorption line from the white dwarf.  

\begin{figure}
  \begin{center}
    \includegraphics[width=\columnwidth]{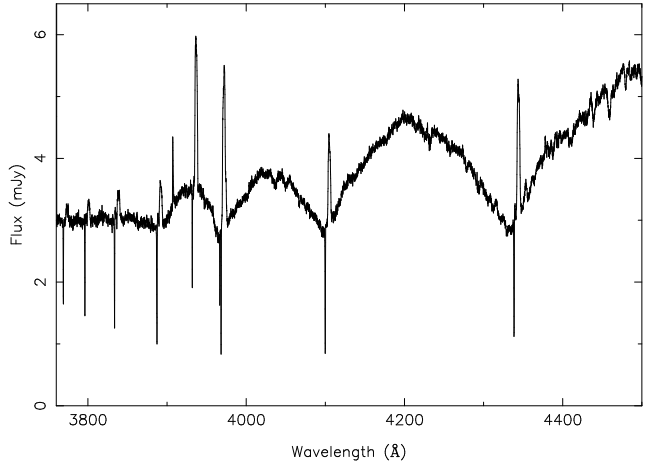}
    \caption{UVES blue arm spectrum of QS\,Vir taken at an orbital phase of
      $\phi=0.22$ showing sharp absorption features from material passing in
      front of the white dwarf. Emission lines are also present originating
      from the M star component caused by a combination of activity and
      irradiation.} 
  \label{fig:prom_spec}
  \end{center}
\end{figure}

In addition to these general features, we also re-detect the narrow absorption
features noted by \citet{parsons11} in the hydrogen Balmer series and calcium
H and K lines, an example of this is shown in Figure~\ref{fig:prom_spec}.
Furthermore, we also detect absorption features in several other species
including the sodium D lines and several \Ion{Mg}{i} and \Ion{Fe}{ii}
lines. The behaviour and origin of these features are discussed in the
following section. 

\subsection{Prominence features}

\begin{figure*}
  \begin{center}
    \includegraphics[width=0.332\textwidth]{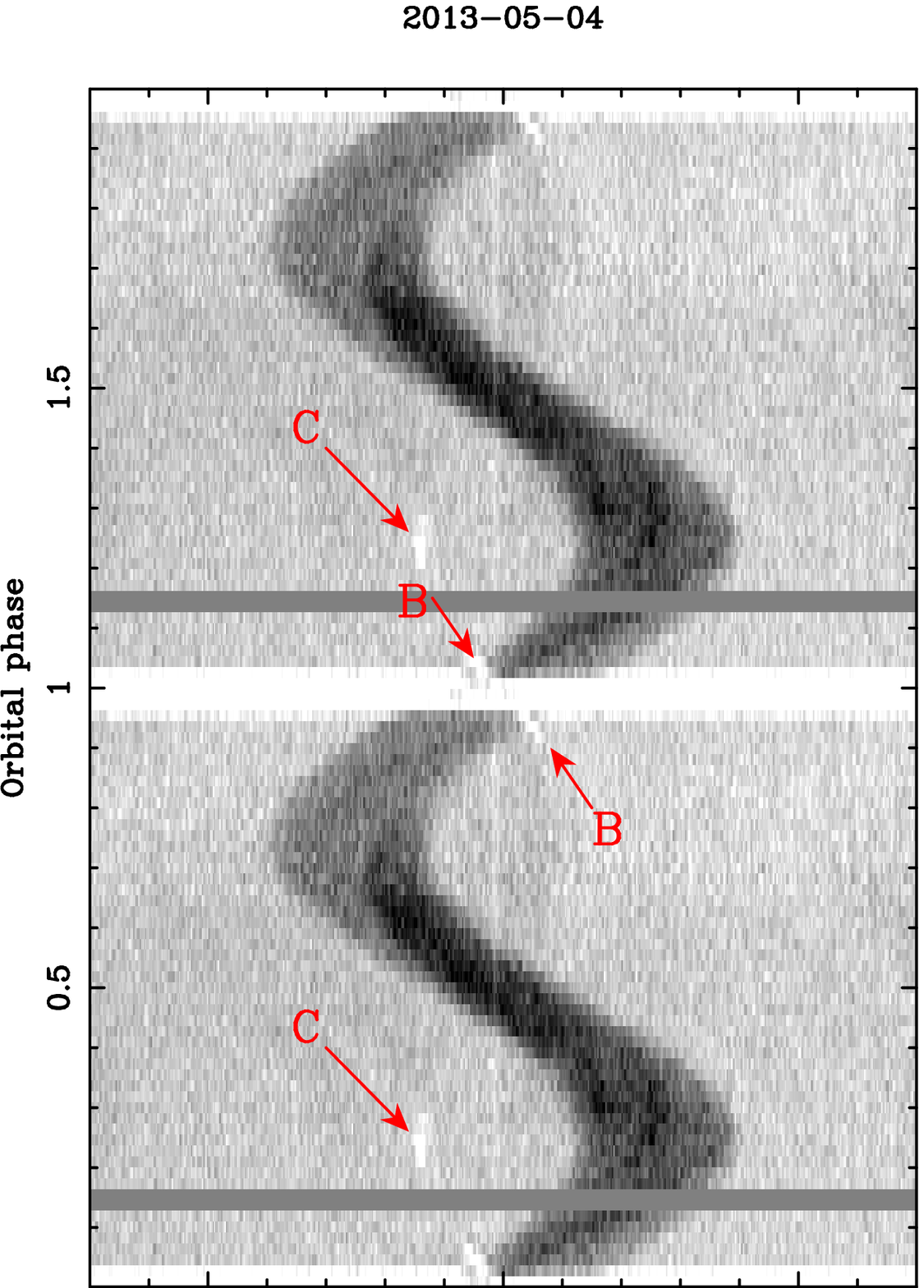}
    \includegraphics[width=0.3\textwidth]{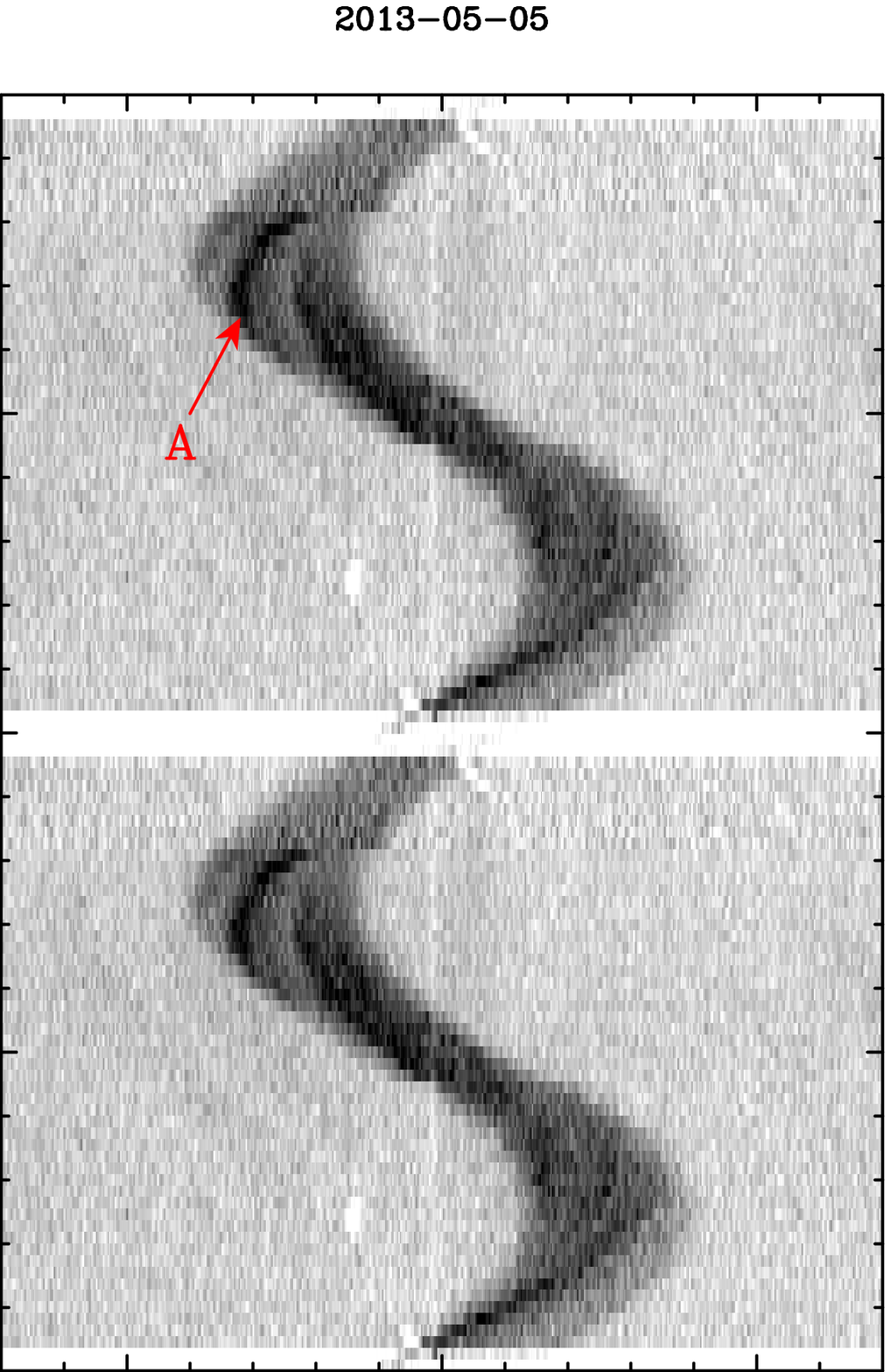}
    \includegraphics[width=0.3\textwidth]{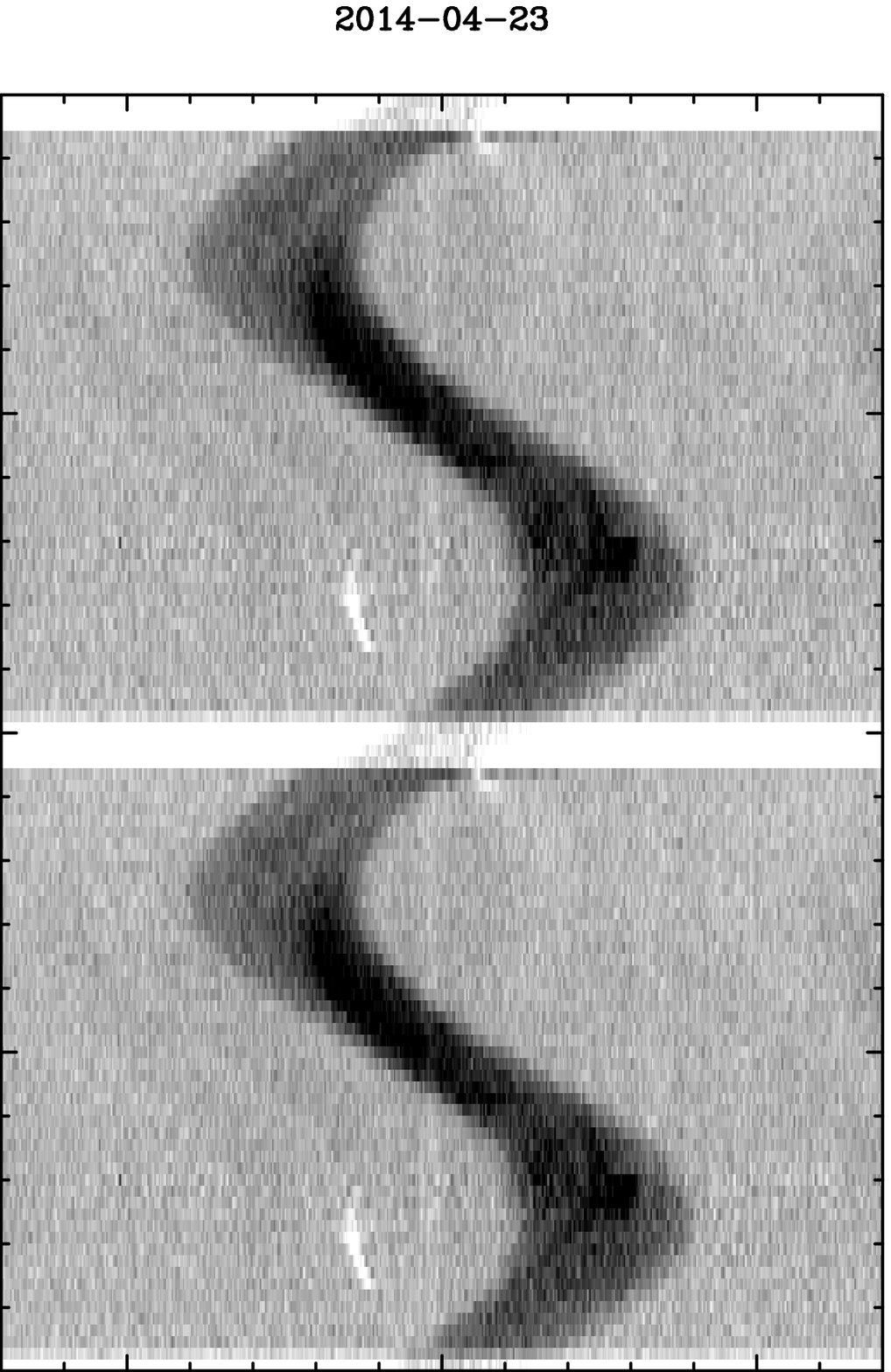}\\
    \vspace{4mm}
    \includegraphics[width=0.332\textwidth]{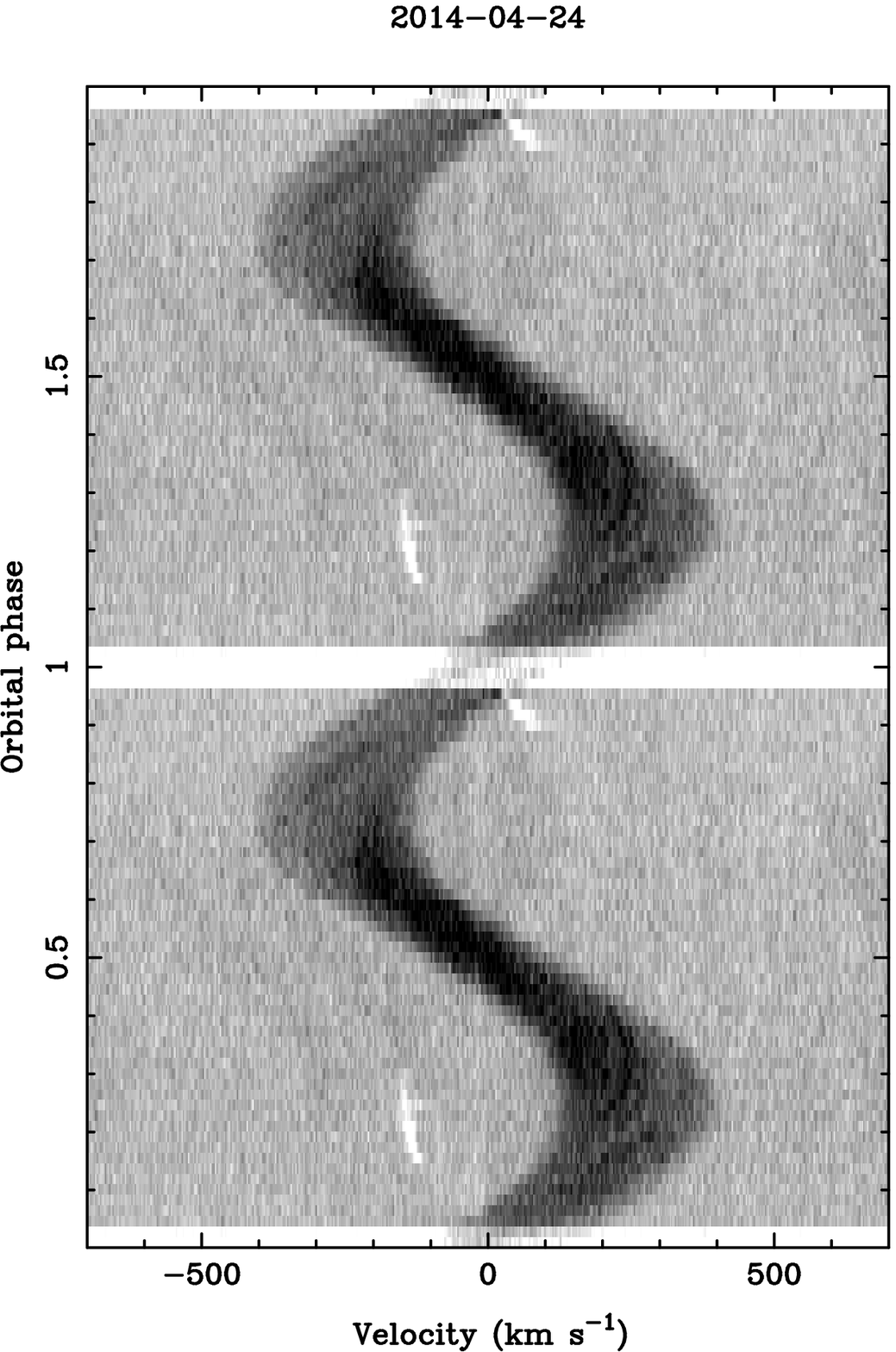}
    \includegraphics[width=0.3\textwidth]{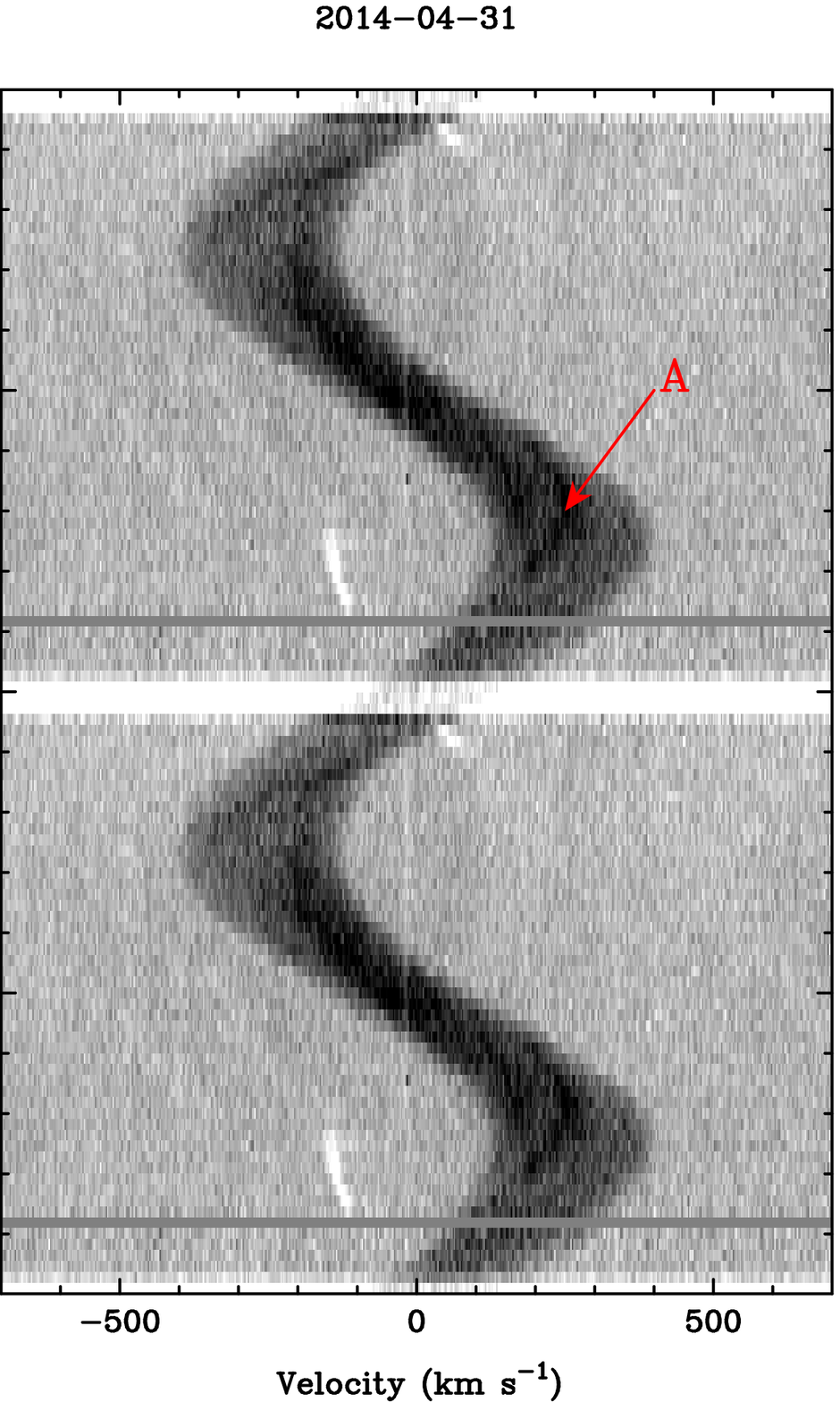}
    \includegraphics[width=0.3\textwidth]{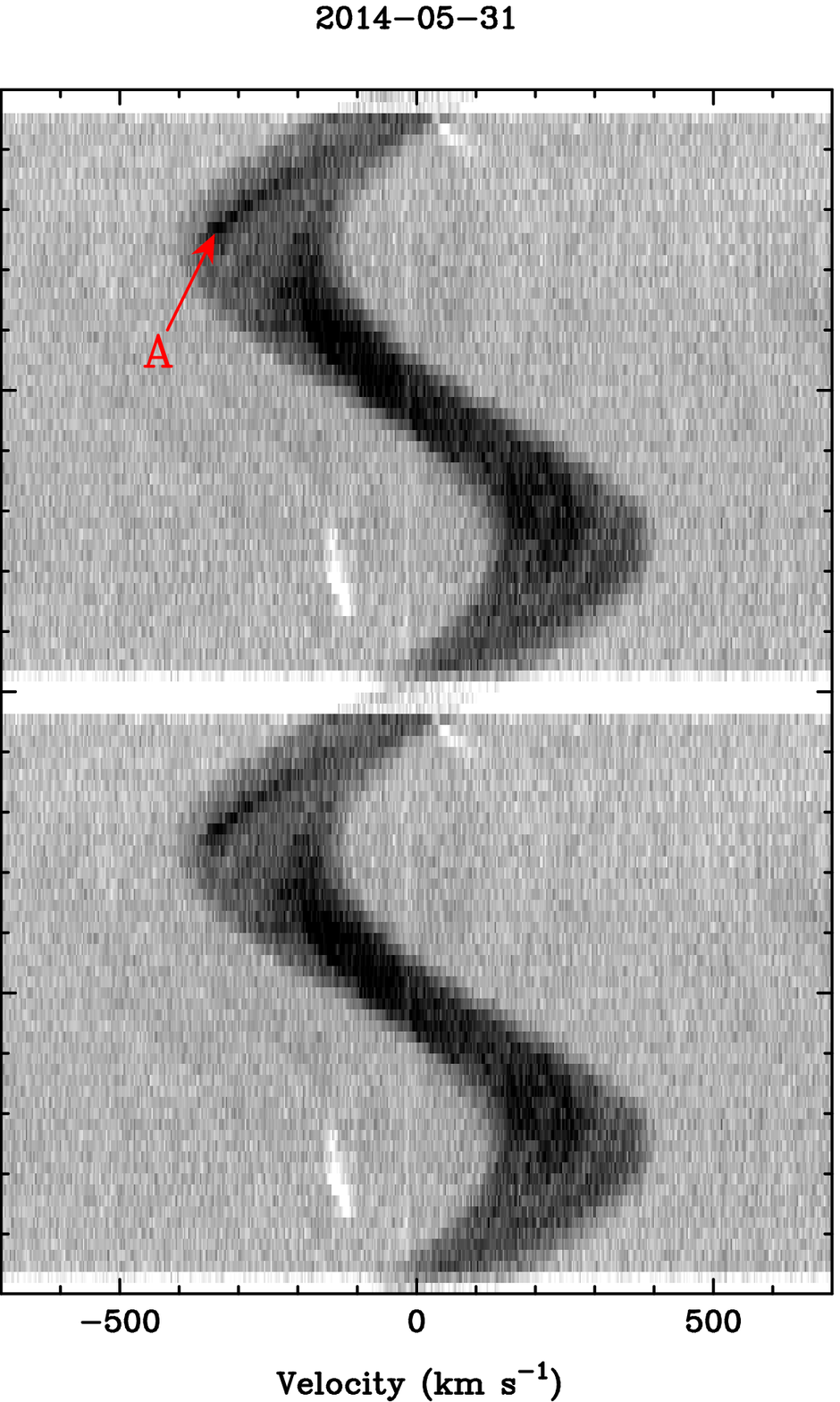}\\
    \caption{Phase-folded trailed spectrograms of the \Ion{Ca}{ii} 3934{\AA}
      line, data shown twice for visual continuity. The date at the start of
      the night the data were obtained is shown above each plot. Higher fluxes
      are darker. The main features visible are the emission from the active M
      star and the eclipse of the white dwarf. However, a narrow absorption
      feature is visible just before and after the eclipse (``B'') and at
      phase 0.25 (``C'') in all the plots. This is caused by material moving
      in front of the white dwarf and, although variable, is always visible,
      despite the data spanning more than a year.} 
    \label{fig:ca_trails}
  \end{center}
\end{figure*}

\begin{figure*}
  \begin{center}
    \includegraphics[width=0.332\textwidth]{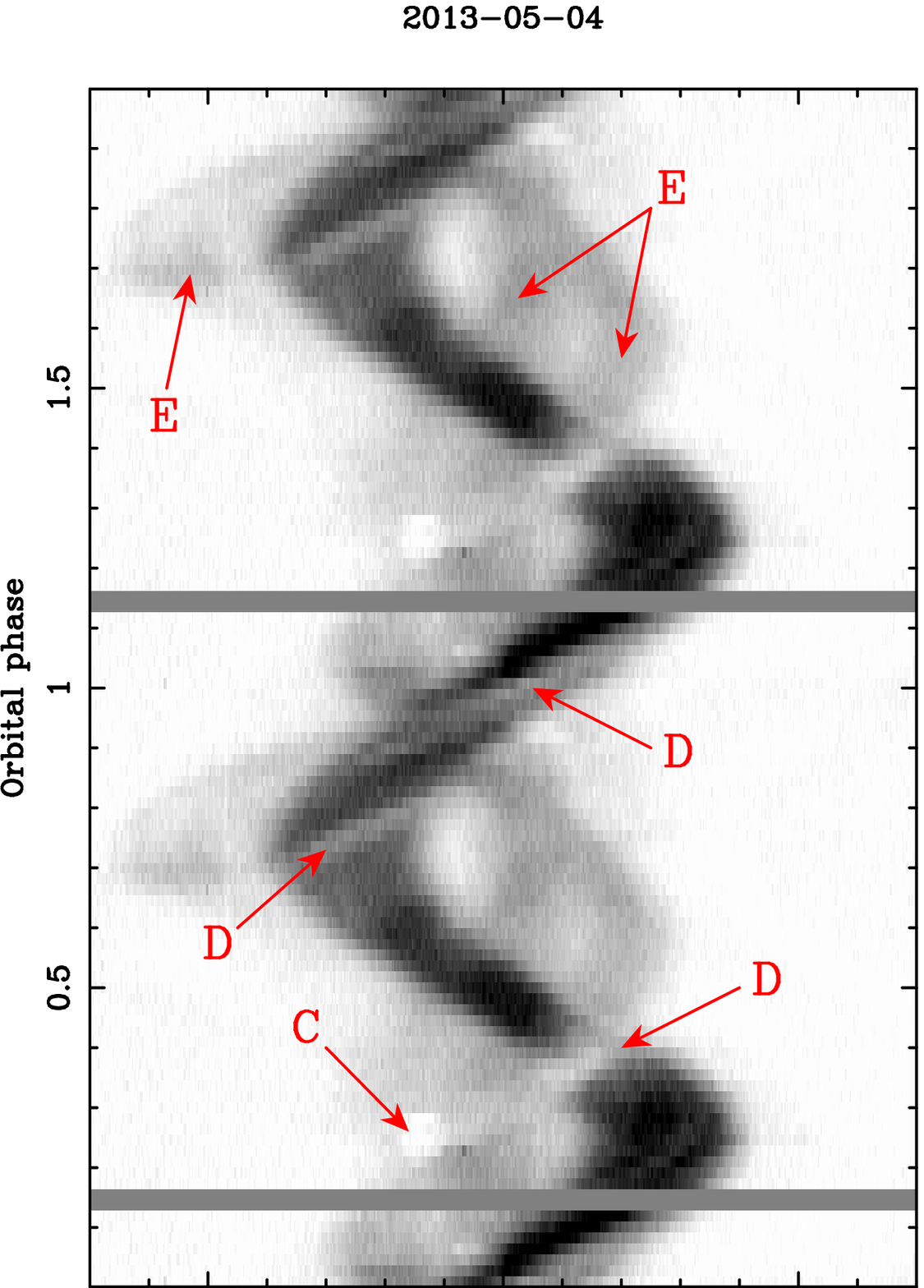}
    \includegraphics[width=0.3\textwidth]{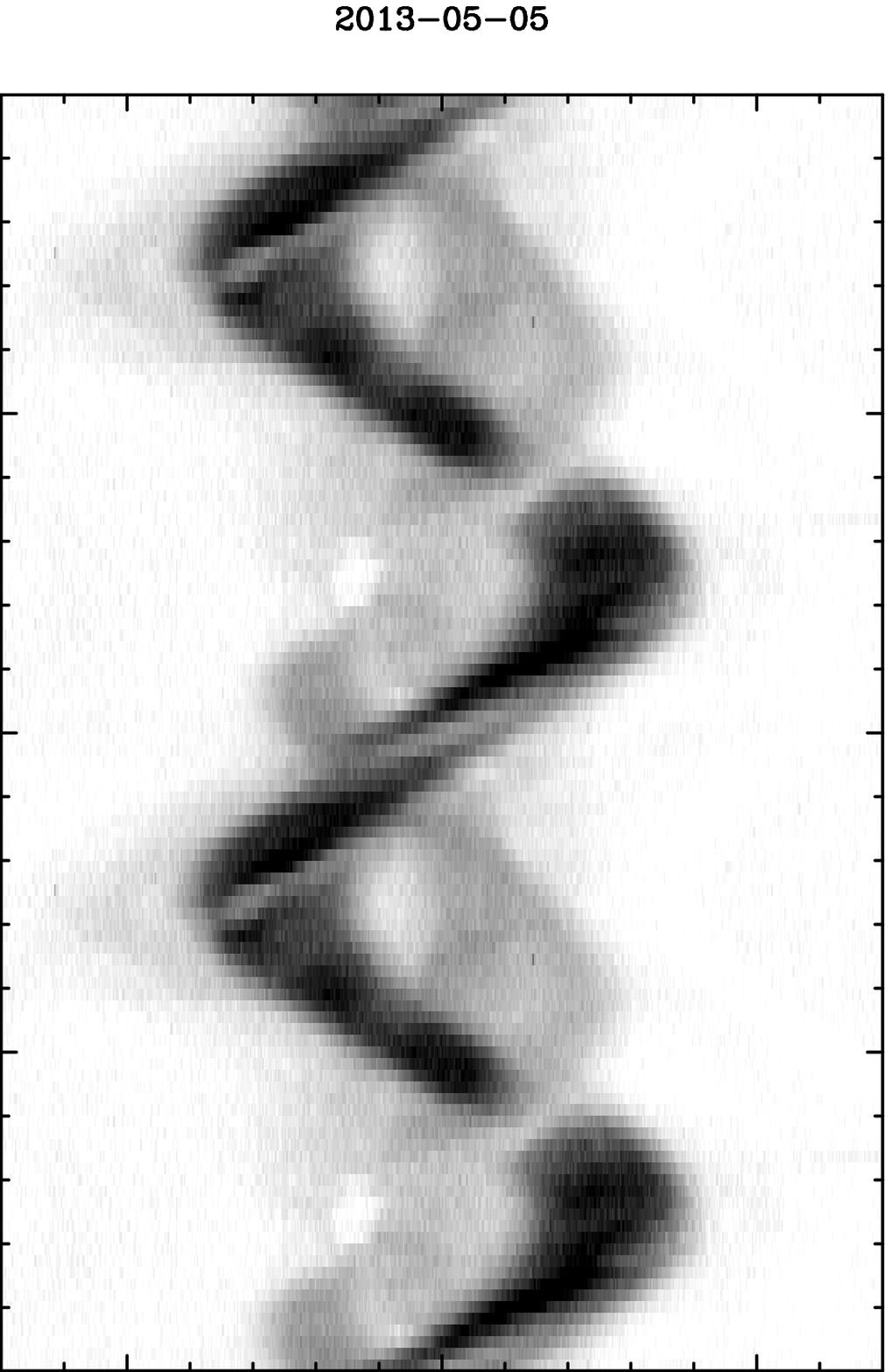}
    \includegraphics[width=0.3\textwidth]{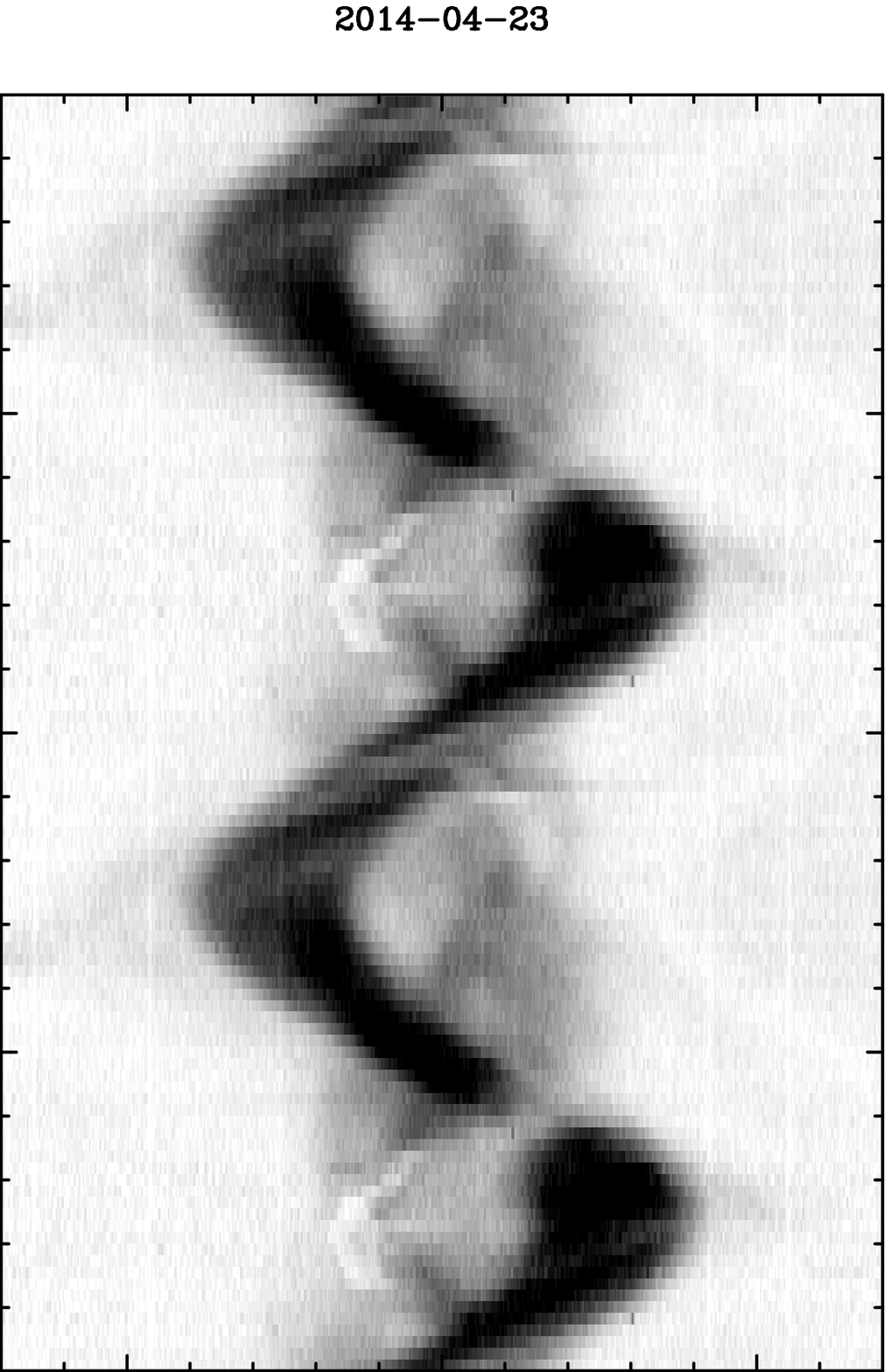}\\
    \vspace{4mm}
    \includegraphics[width=0.332\textwidth]{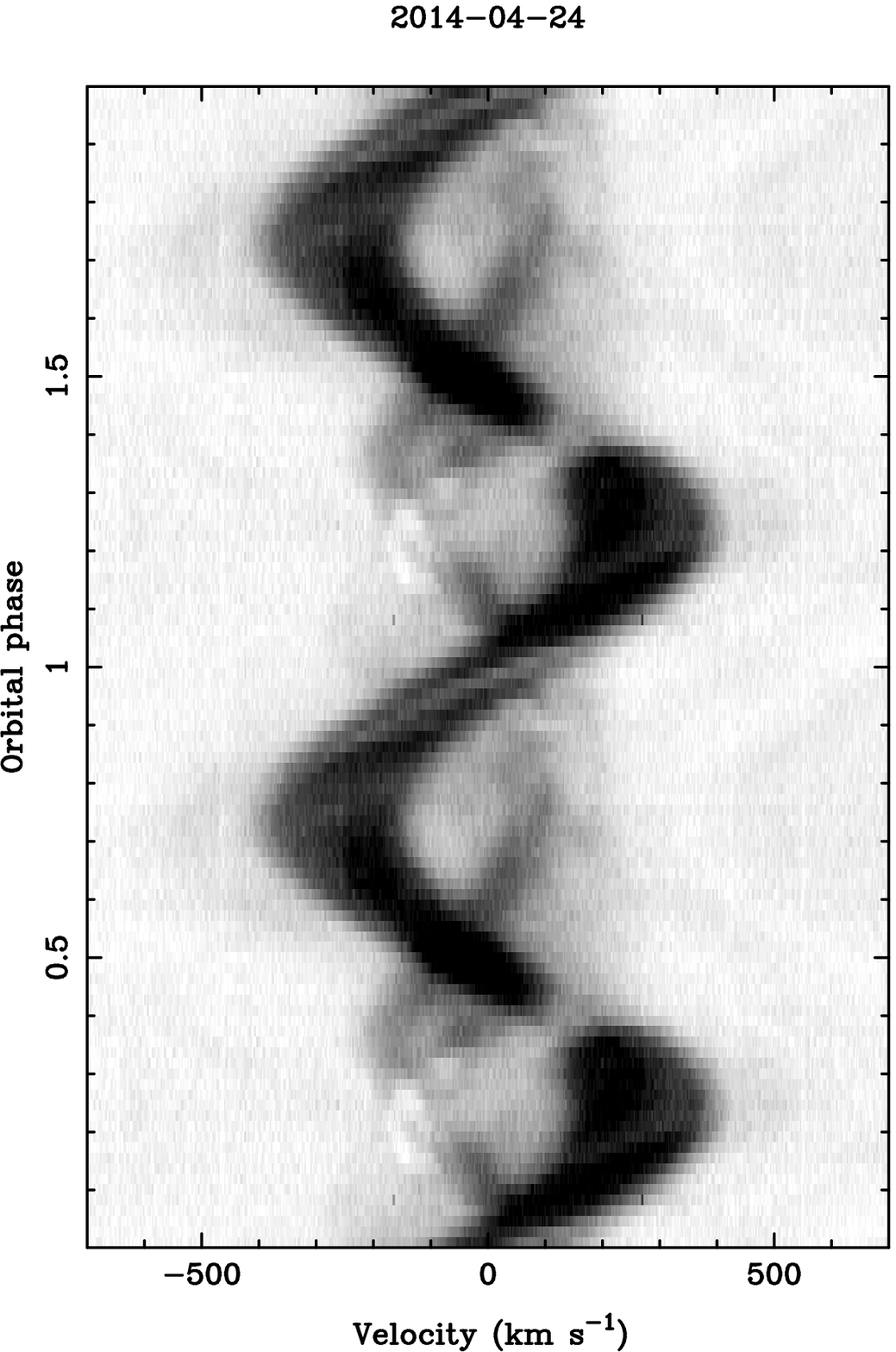}
    \includegraphics[width=0.3\textwidth]{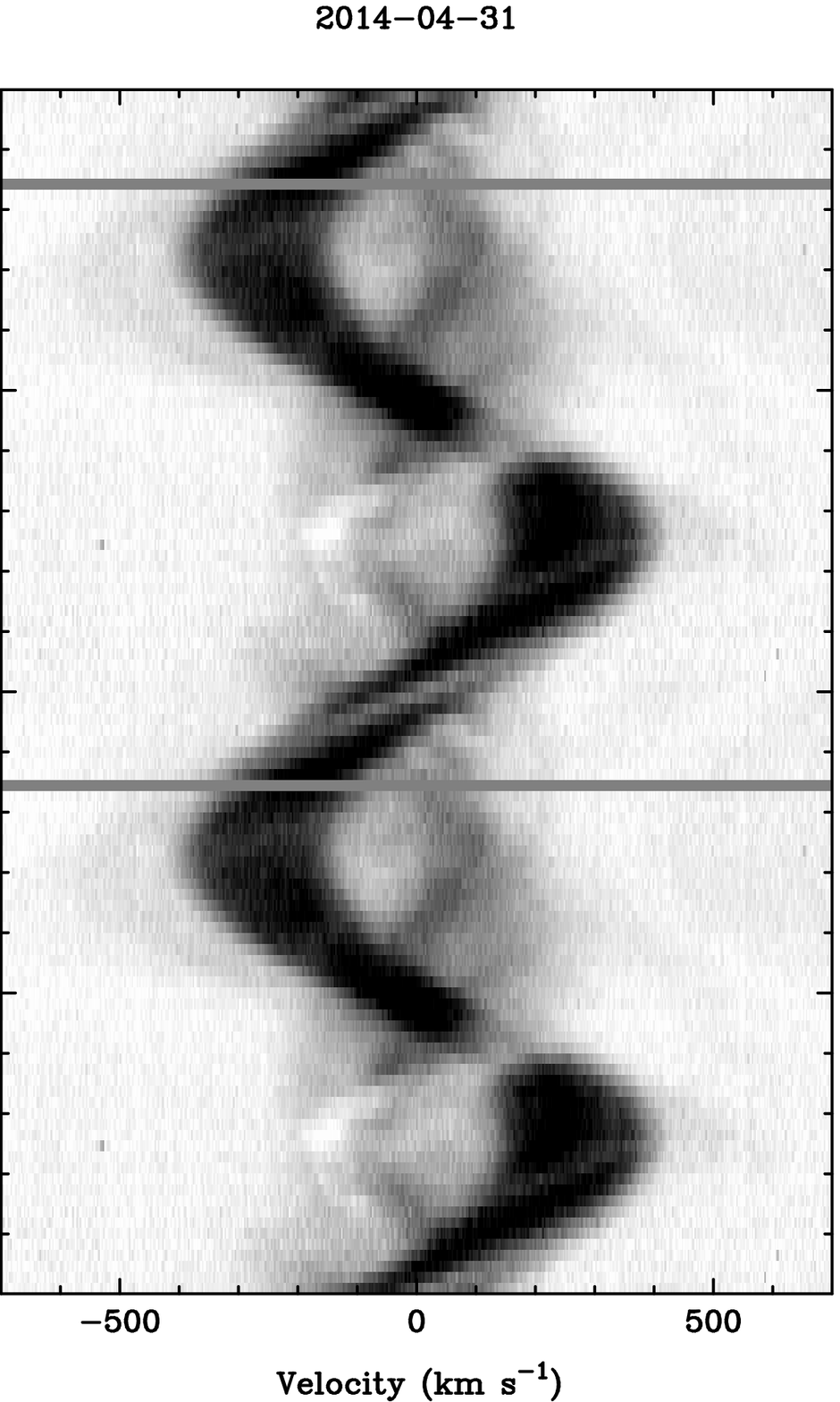}
    \includegraphics[width=0.3\textwidth]{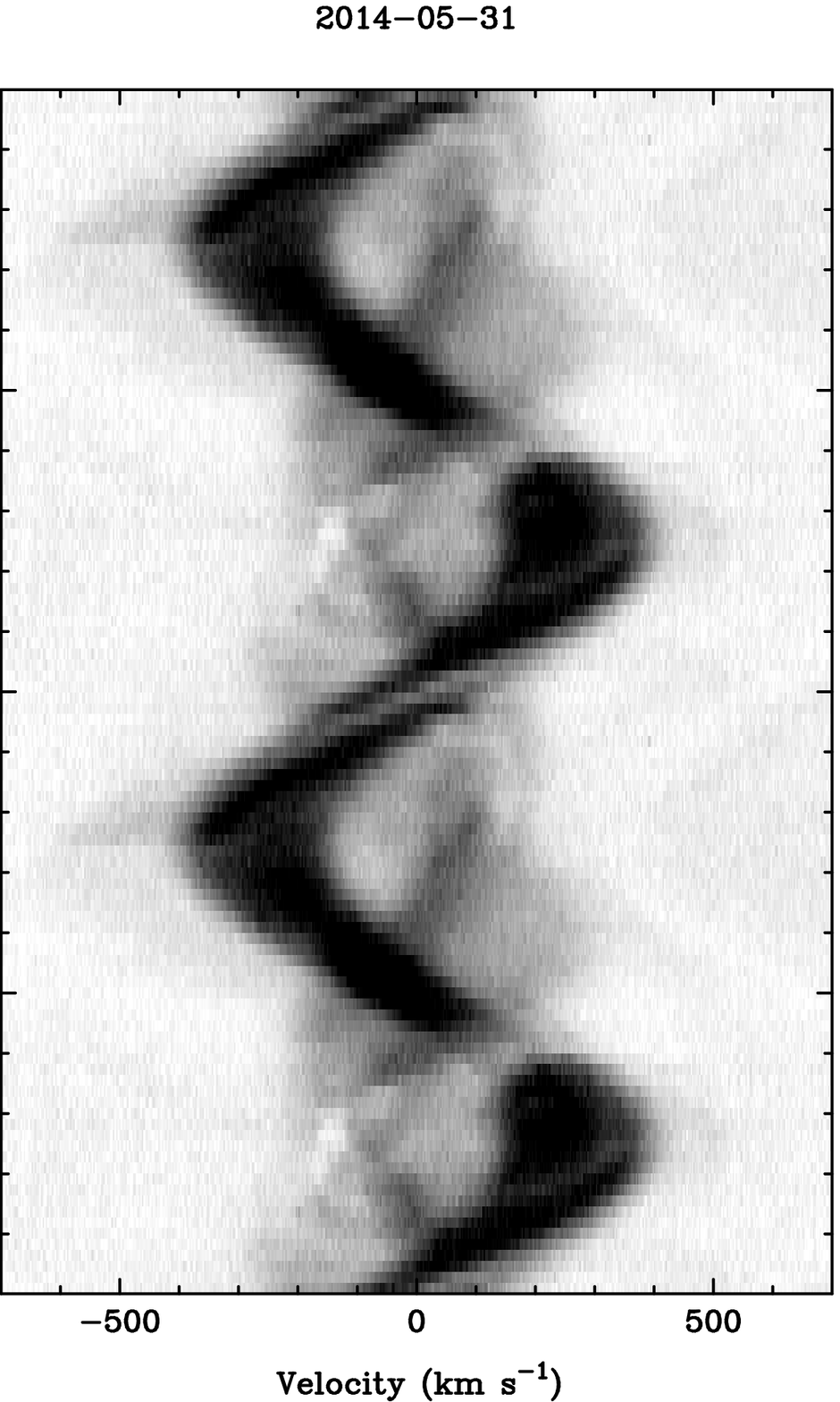}\\
    \caption{Phase-folded trailed spectrograms of the H$\alpha$ line, data
      shown twice for visual continuity. The continuum has been subtracted
      from each spectrum and the date at the start of the night the data were
      obtained is shown above each plot. Higher fluxes are darker. The
      strongest component is emission from the active M star. However, there
      are at least two other emission components (``E''), neither of which
      originate from either of the stars. Material is seen moving across the
      face of both the white dwarf (``C'') and M star (``D''). Despite being
      separated by more than a year, the trails look remarkably similar.} 
    \label{fig:ha_trails}
  \end{center}
\end{figure*}

The sharp absorption features detected by \citet{parsons11} were attributed to
a prominence from the M star passing in front of the white dwarf. However, only
a single spectrum was available, limiting the constraints that could be placed
on its location and size. Our new data cover the full binary orbit and hence
much tighter constraints can be placed. Moreover, our data cover multiple
orbits over a timescale of more than a year, allowing us to investigate the
evolution of these features. Figure~\ref{fig:ca_trails} shows trailed
spectrograms of the \Ion{Ca}{ii} 3934{\AA} line on 6 separate nights, covering
timescales from a day to a year. Strong emission from the M star is visible
resulting from a combination of activity and irradiation (note how the
emission is strongest at phase 0.5, when the irradiated face points towards
us). There are additional short-lived localised emission components from the M
star labelled ``A'' in Figure~\ref{fig:ca_trails}, best seen in the 5th of May
2013 data (middle-top panel of Figure~\ref{fig:ca_trails}). These narrow
emission features occur on top of the intrinsic emission from the M star and
only persist for single orbits, they are likely related to flares on its
surface. There is also a narrow non-variable, zero velocity absorption
component likely the result of interstellar absorption.

In addition to emission from the M star, there is also a narrow absorption
line moving in anti-phase to the M star, but only visible at certain orbital
phases. This absorption is seen just before and after the eclipse of the white
dwarf (labelled ``B'' in Figure~\ref{fig:ca_trails}, particularly clear in the
2013 data) as well as close to phase 0.25, labelled ``C''. When visible, these
absorption features have the same velocity as the white dwarf (see
Section~\ref{sec:rvwd}), implying that this material blocking the light of the
white dwarf has no radial velocity relative to it, and suggests that it is in
a state of solid-body rotation with respect to the binary, likely in the form
of prominences.

The most striking aspect of Figure~\ref{fig:ca_trails} is that these
prominence features appear to persist on timescales of days, weeks, months and
years. Given the similarity of the prominence feature detected at on orbital
phase of 0.16 in the 2002 spectra by \citet{parsons11}, it also appears
that these features could be stable on timescales of decades. However,
Figure~\ref{fig:ca_trails} also shows that they are not completely fixed. The
absorption feature around phase 0.25 is visible for longer during 2014
compared to the 2013 data. Meanwhile the absorption around the eclipse is 
much clearer in the 2013 data and is only obvious before the eclipse in the
2014 dataset.

In Figure~\ref{fig:ha_trails} we show the same plots as
Figure~\ref{fig:ca_trails} but for the H$\alpha$ line. This line shows a much
more complex behaviour, with multiple absorption and emission components
visible. The strongest component is still the emission from the M star, but
narrow absorption components are seen crossing this emission throughout the
orbit during all six observations (the strongest are labelled ``D'' in
Figure~\ref{fig:ha_trails}). The prominence material passing in front of 
the white dwarf is still visible (``C''), albeit slightly weaker than in the
\Ion{Ca}{ii} line since the white dwarf contributes a smaller 
amount of the overall flux at H$\alpha$. Furthermore, there are several
additional emission components which do not correspond to either star in the
binary, labelled ``E''. The H$\alpha$ line also shows considerable variability
in the strength of all these features from night-to-night compared to the
\Ion{Ca}{ii} line.

\begin{figure*}
  \begin{center}
    \includegraphics[width=\columnwidth]{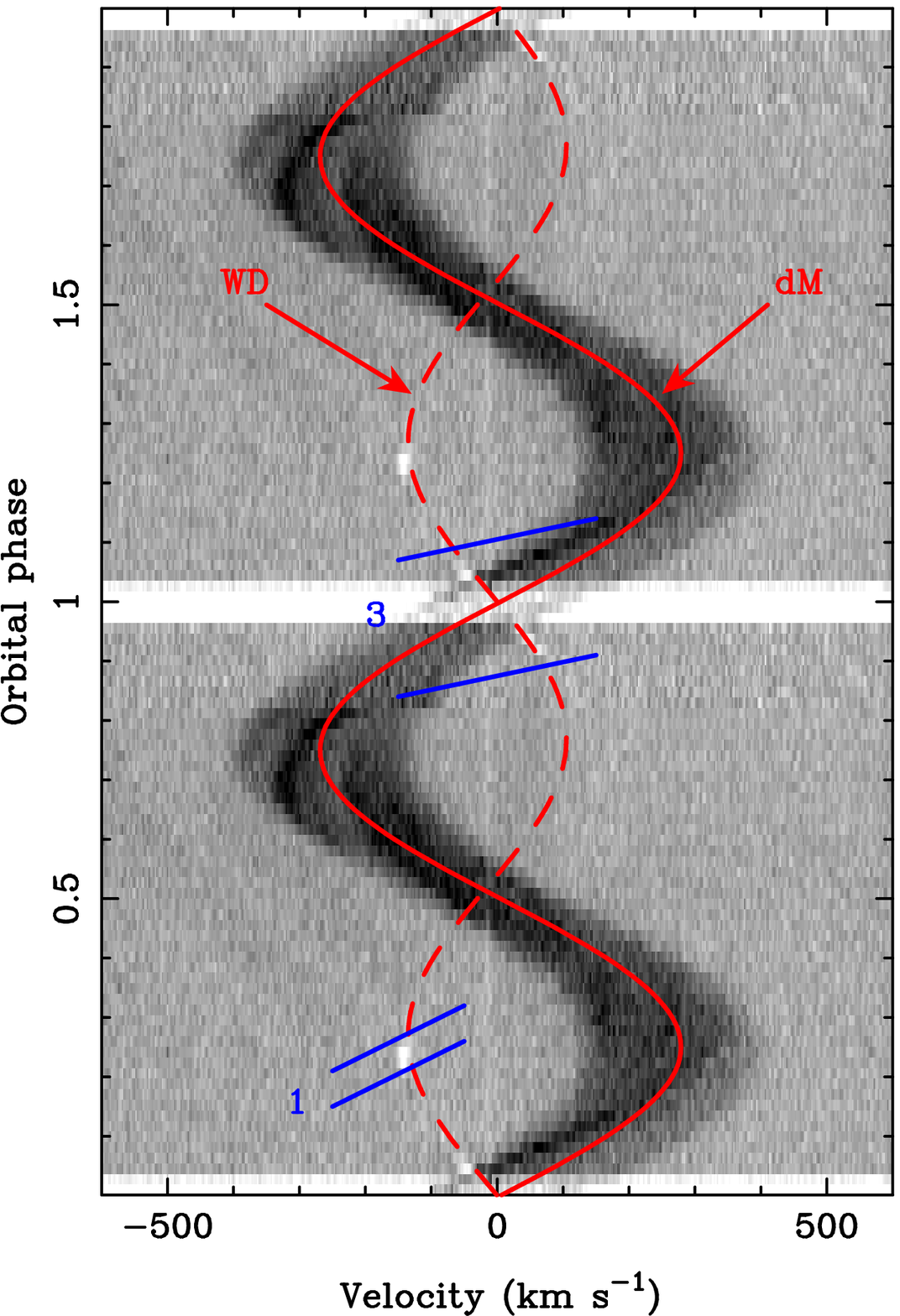}
    \includegraphics[width=0.9417\columnwidth]{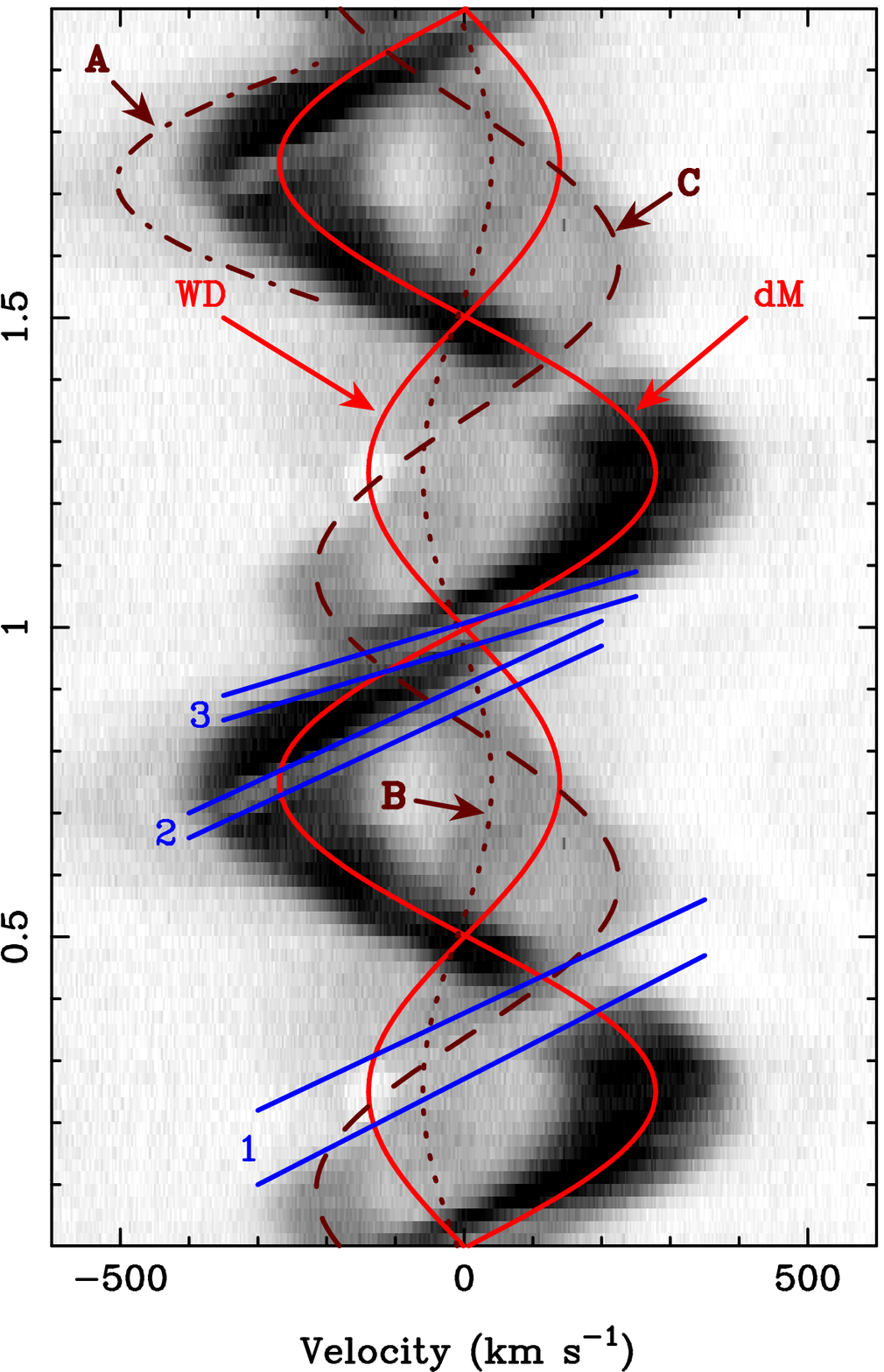}
    \caption{Trailed spectrogram of the \Ion{Ca}{ii} 3934{\AA} (left) and
      H$\alpha$ (right) lines from 5th May 2013 with the main features
      labelled. The velocities of the two stars are indicated by the bright red
      sinusoids. In blue we highlight three clear absorption features that
      cross one of both of the stars (only two are visible in the \Ion{Ca}{ii}
      trail). We also highlight three other emission components in the
      H$\alpha$ trail with dark red lines.} 
  \label{fig:emis_comps}
  \end{center}
\end{figure*}

In Figure~\ref{fig:emis_comps} we show a larger version of the \Ion{Ca}{ii}
and H$\alpha$ trailed spectrogram from the second observing run (5th of May
2013) with some of the major features labelled. The radial velocity of the
white dwarf and M star are shown as red lines (see Section~\ref{sec:rvwd} and
\ref{sec:rvmd}). We highlight in blue three clear prominence-like absorption
features visible in all six observations, that appear to cross one or both of
the stars. In addition we highlight three additional emission components
visible in the H$\alpha$ trail that do not originate from either star.

\begin{figure}
  \begin{center}
    \includegraphics[width=\columnwidth]{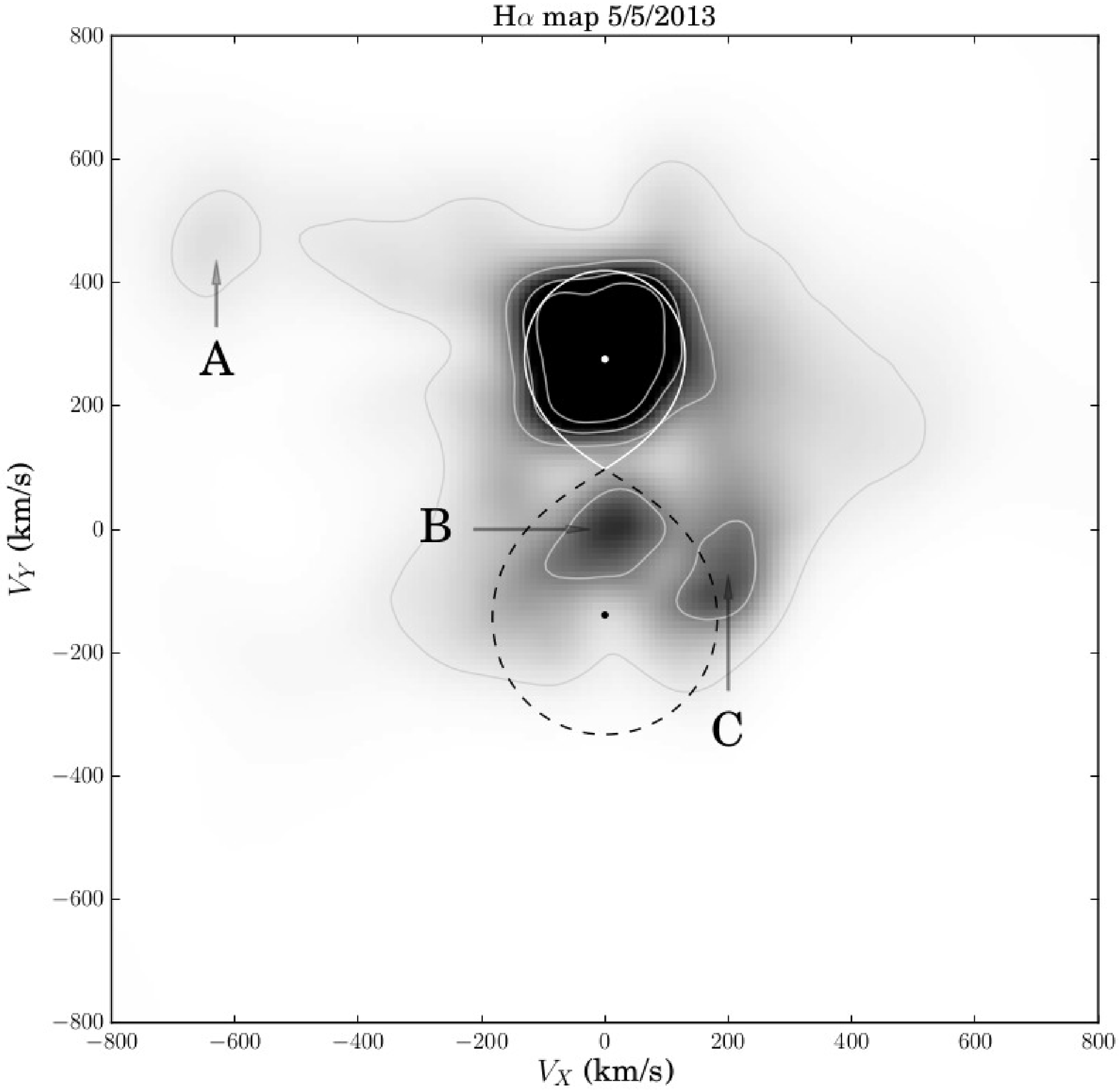}
    \caption{Doppler map of the H$\alpha$ line from 5th May 2013. The Roche
      lobes of the white dwarf and M star are highlighted with a black dashed
      line and solid white line respectively. To highlight the additional
      emission features against the strong emission from the M star we have
      also plotted contours. Three clear emission features that do not
      originate from either star are labelled.} 
  \label{fig:dmap}
  \end{center}
\end{figure}

To further highlight the various emission components visible in the trailed
spectrogram of the H$\alpha$ line we used Doppler tomography \citep{marsh88}.
We used the MODMAP code described in \citet{steeghs03} to reconstruct the
Doppler map shown in Figure~\ref{fig:dmap}, using all the data from May 5th
2013. It clearly shows that the strongest emission component originates from
the M star itself, but the three additional emission components visible in the
trailed spectrogram are seen more clearly, with two close to the white dwarf's
Roche lobe and one high velocity emission component. These components are
labelled ``A'', ``B'' and ``C'' and correspond to the additional emission seen
in the H$\alpha$ trail in Figure~\ref{fig:emis_comps}, where we have used the
same labels to highlight the features.

The clearest prominence feature in Figure~\ref{fig:emis_comps}, labelled ``1'',
is responsible for the narrow absorption line seen at around phase 0.25 in the
\Ion{Ca}{ii} line, when it passes in front of the white dwarf. At around phase
0.4 this same material passes in front of the M star, completely blocking the
H$\alpha$ emission component from this star. Furthermore, this prominence is
seen in emission at certain orbital phases, highlighted as component
  ``C'' in both Figure~\ref{fig:emis_comps} and Figure~\ref{fig:dmap}. It is
possible that the prominence is dense enough that it is able to reprocess the
light it receives from the white dwarf. Alternatively, this could be intrinsic
emission from the prominence. Figure~\ref{fig:ha_trails} shows that the
strength of this emission clearly varied from 2013 to 2014, weakening in 2014.

\begin{figure}
  \begin{center}
    \includegraphics[width=0.95\columnwidth]{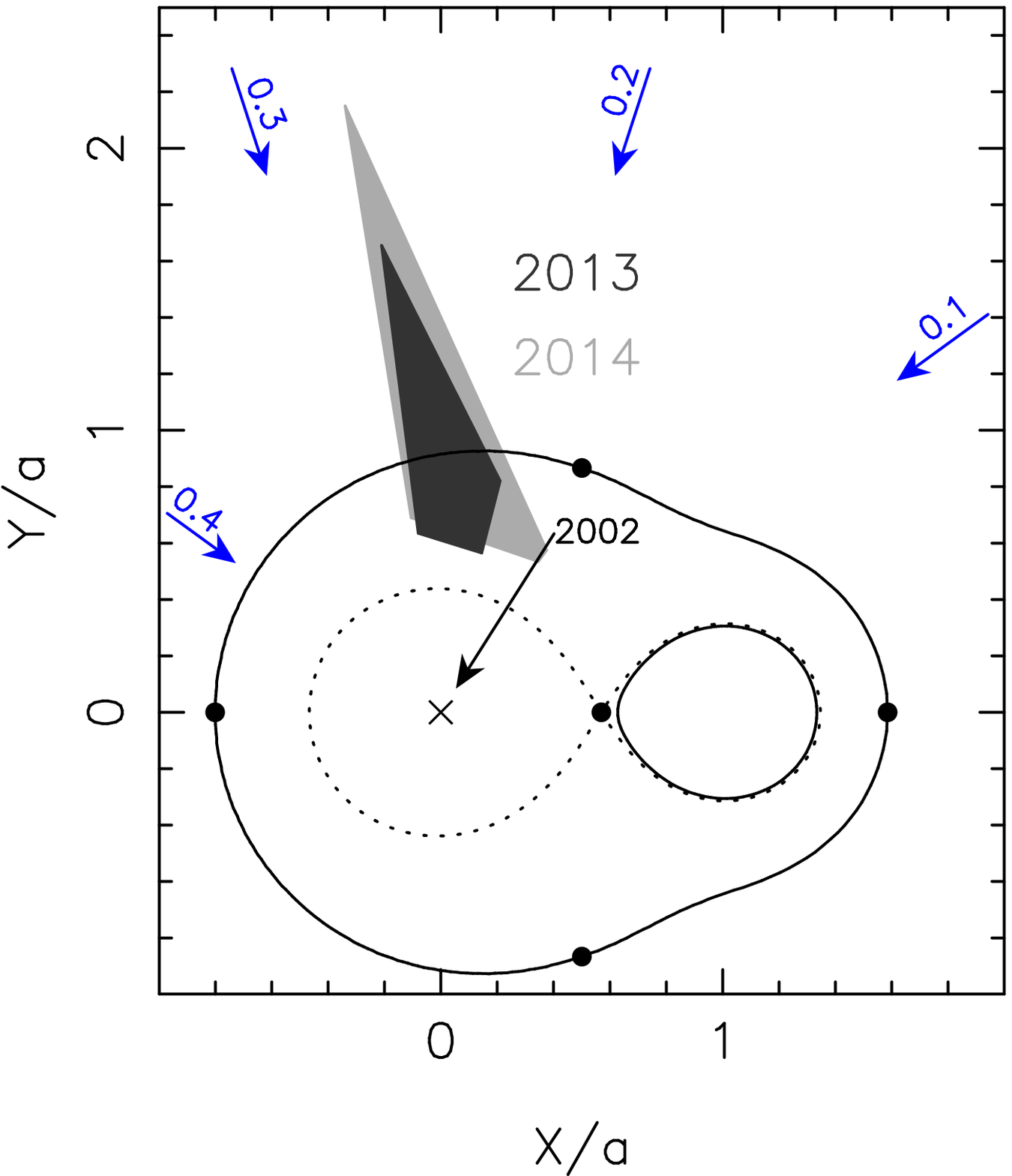}
    \caption{A top down view of the binary indicating the location of the
      large prominence feature in 2013 and 2014 (feature ``1'' in
      Figure~\ref{fig:emis_comps}), the stars rotate clockwise and the viewing
      angle to the white dwarf at various orbital phases is shown in blue. The
      dotted lines indicate the Roche lobes of the two stars, the radius of
      the M star is shown as a solid line. The arrow indicates the viewing
      angle to the white dwarf during the 2002 observation from
      \citet{parsons11} in which prominence material was also observed. The
      cross marks the location of the white dwarf and the black points
      indicate the location of the five Lagrange points. The outer solid black
      line indicates the effective ``co-rotation'' radius of the binary (see
      the text for a definition of this radius).}  
  \label{fig:visual}
  \end{center}
\end{figure}

The fact that this prominence crosses in front of both stars allows us to
triangulate its position within the binary. Figure~\ref{fig:visual} shows a
visualisation of the binary indicating both the location and size of this
prominence. We show this for both 2013 and 2014, variations on shorter
timescales are minimal. As is evident from Figures~\ref{fig:ca_trails} and
\ref{fig:ha_trails} the prominence is larger in 2014, clearly taking longer to
pass in front of the white dwarf. We also show the viewing angle to the white
dwarf in the 2002 data from \citet{parsons11} which also passed through a
similar prominence feature.

The stability and longevity of this feature is remarkable, given
that it is located at an unstable location within the binary and hence the
material should be expelled quickly. In Figure~\ref{fig:visual} we plot the
effective ``co-rotation'' radius of the binary. While this radius is well
defined for a single star (the radius at which the gravitational attraction
equals the centripetal acceleration of a particle in solid-body rotation with
the star, hence a particle can sit there with no other force needed to hold
it), such a radius does not exist for a binary. In this case the centripetal
acceleration acts towards the centre of mass, but the gravitational forces act
towards the two stars and so there is not a stable radius, but rather five
equilibrium  points (the Lagrangian points). However, in
Figure~\ref{fig:visual} we slightly extend the definition by considering the
radial component of acceleration in the rotating frame (automatically
including gravity and centripetal terms) as measured from the centre of mass
of the M star, resulting in a line that passes through the Lagrangian
points. Nevertheless, except for at the Lagrangian points themselves, there
are still forces acting perpendicular to this line, so it does not represent a
stable region within the binary, hence some additional force is required to
keep material in this region. However, it may well be easier for material to
stay fixed near this line than at other locations. From
Figure~\ref{fig:visual} it is evident that material is located at this radius,
but also extends substantially beyond it, requiring that quite significant
non-gravitational forces are acting upon it (presumably some sort of magnetic
force). 

The feature labelled ``2'' in Figure~\ref{fig:emis_comps} crosses the M star at
an orbital phase of $\sim$0.75, but does not appear to pass in front of the
white dwarf (hence it is not visible in the \Ion{Ca}{ii} trail), meaning that
we cannot completely determine its location or size like the large
prominence. It is possible that it passes in front of the white dwarf shortly
before or during the eclipse, in which case it would be located close to the
surface of the M star. Or instead, this particular prominence could be at an
inclination such that it only passes in front of the M star. This prominence
weakened considerably between 2013 and 2014 and is barely visible in the last
observations in Figure~\ref{fig:ha_trails}. There does not appear to be an
obvious emission component associated with this prominence.

The feature labelled ``3'' in Figure~\ref{fig:emis_comps} crosses the face of
the M star during the eclipse of the white dwarf, and therefore this material
must be located behind the M star, from the white dwarf's
perspective. Moreover, the \Ion{Ca}{ii} trail shows that this material
also passes in front of the white dwarf just before and after the
eclipse. Unfortunately, this means that we cannot determine the exact location
of this material either. This prominence is seen in emission as a high
velocity component blueward of the M star at phase 0.75 (labelled ``A'') in
the H$\alpha$ trail (dot-dashed line) and more clearly at (-650,450) in
Figure~\ref{fig:dmap}. Similar absorption features were detected in
ultraviolet spectra of V471\,Tau just before and after the eclipse of the
white dwarf \citep{guinan86}, which they attributed to cool coronal
loops. Interestingly, \citet{odonoghue03} detected a sharp pre-eclipse dip in
the light curve of QS\,Vir (see their Figure~8) which could be from similar
coronal loops, implying that this material can sometimes be dense enough to
block out continuum light. However, the lack of a comparison star during this
observation means that this dip should be interpreted with some
caution. Despite our extensive photometric monitoring
(e.g. Table~\ref{tab:obslog}) this dip has not been seen since.

Finally, there is another emission component highlighted in the H$\alpha$
trail in Figure~\ref{fig:emis_comps} by a dotted line (labelled ``B''). It
moves in anti-phase with the M star, but with a much lower velocity than the
white dwarf, meaning that this must originate from material between the two
stars. It is also visible in Figure~\ref{fig:dmap} close to (0,0), within the
Roche lobe of the white dwarf. Figure~\ref{fig:ha_trails} shows that this
emission was much stronger in 2014 compared to 2013. Assuming that this
material co-rotates with the binary, its location is given by  
\begin{equation}
\frac{K_\mathrm{mat}}{K_\mathrm{sec}} = (1+q)\left(\frac{R}{a}\right) - q,
\end{equation}
where $K_\mathrm{mat}$ and $K_\mathrm{sec}$ are the radial velocity
semi-amplitudes of the material and the M star respectively, $q$ is the mass
ratio and $R/a$ is the distance of the material from the white dwarf scaled by
the orbital separation ($a$).

Using our measured values of $K_\mathrm{sec}$ and $q$ (see
Sections~\ref{sec:roche_tom} and \ref{sec:lc_model}) and
$K_\mathrm{mat}=-100$\kms determined by Gaussian fitting in the same way as
outlined in the next section, gives $R/a=0.09$. Similar emission features have
been seen in other post common-envelope binaries
\citep{parsons13mag,tappert11} as well as CVs
\citep{steeghs96,gansicke98,kafka05,kafka06} and could be related to another 
prominence structure or heated material from the wind of the M star close to
the white dwarf. Interestingly, there appears to be a genuine deficiency
  of H$\alpha$ emission from the inner face of the M star, best seen in
  Figure~\ref{fig:dmap}. The deficiency may be due to this material between
  the two stars shielding the M star from irradiation by the white dwarf.

Despite being an eclipsing system, the inclination of QS\,Vir is relatively
low ($i=77.7^\circ$, see Section~\ref{sec:lc_model}), therefore, not only are
these prominences far from the M star, they are also located far from its
equatorial plane. However, many cool, isolated low-mass stars show similar
prominence systems beyond their co-rotation radii \citep{donati99,dunstone06},
and located far from their equatorial planes \citep{collier89}. Indeed,
\citet{jardine01} found that for single stars there exist stable locations
both inside and outside the co-rotation radius and up to 5 stellar radii from
the star \citep{jardine05}. Long-lived prominences have also been detected in
the CVs BV\,Cen \citep{watson07}, IP\,Peg and SS\,Cyg \citep{steeghs96} and
AM\,Her \citep{gansicke98}. Magnetically confined material was also detected
in BB\,Dor \citep{schmidtobreick12} and near the L4 and L5 Lagrange points in
AM\,Her \citep{kafka08}. In addition, the detached white dwarf plus
main-sequence star binaries V471\,Tau \citep{jensen86} and SDSS\,J1021+1744
\citep{irawati15} both show long-lived prominence-like features, in very
similar positions to the large prominence in QS\,Vir.

\subsection{White dwarf radial velocity amplitude} \label{sec:rvwd}

We measure the radial velocity amplitude of the white dwarf using the narrow
\Ion{Mg}{ii} 4481{\AA} absorption line. Due to emission from the M star, the
cores of the hydrogen Balmer lines are not visible, thus making them
unsuitable for radial velocity work. No other intrinsic white dwarf absorption
features are detected in the UVES spectra.

\begin{figure}
  \begin{center}
    \includegraphics[width=\columnwidth]{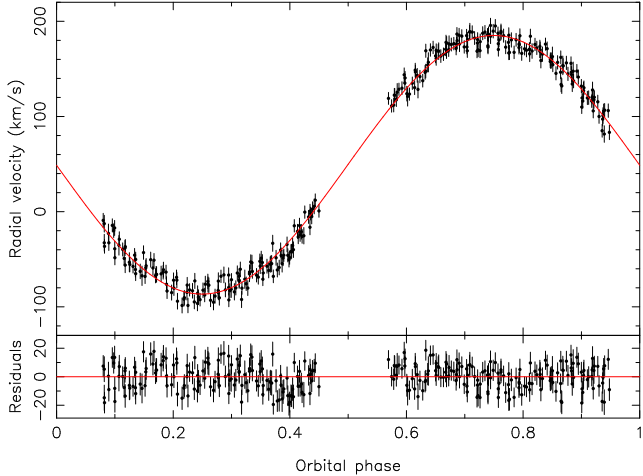}
    \caption{Radial velocity measurements of the \Ion{Mg}{ii} 4481{\AA} absorption line from the white dwarf with fit (red line). The residuals of the fit are shown in the bottom panel.}
  \label{fig:wd_rv}
  \end{center}
\end{figure}

We fit the \Ion{Mg}{ii} line in each spectrum with a combination of a straight
line and a Gaussian component. We do not fit the spectra obtained whilst the
white dwarf was in eclipse (including the ingress and egress
phases). Furthermore, we do not fit the spectra taken during phases
0.45--0.55, since during these phases \Ion{Mg}{ii} emission from the heated
inner hemisphere of the M star fills in the absorption as the two components
cross over. The resultant velocity measurements and their corresponding orbital
phases were then fitted with a sinusoid to determine the white dwarf's radial
velocity semi-amplitude. The result of this is shown in
Figure~\ref{fig:wd_rv}. We find a radial velocity semi-amplitude for the white
dwarf of $139.0\pm1.0$\kms, with a mean velocity of $43.2\pm0.8$\kms,
consistent with (but much more precise than) the measurement from
\citet{odonoghue03} ($137\pm10$\kms and $61\pm10${\kms} respectively) using
{\it  Hubble Space Telescope} (HST) spectra, although our mean velocity is a
little lower than their measured value. The mean velocity of the white dwarf
is redshifted from the systemic velocity of the binary due to its high
gravity, which can be used to constrain its physical parameters (see
Section~\ref{sec:lc_model}).

\subsection{M star radial velocity amplitude} \label{sec:rvmd}

\begin{figure}
  \begin{center}
    \includegraphics[width=\columnwidth]{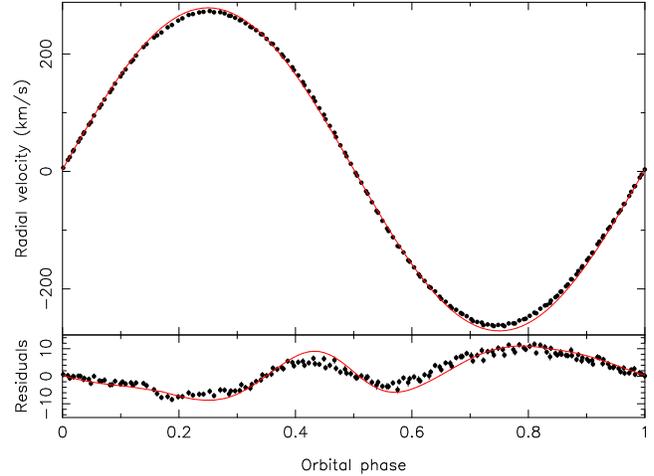}
    \caption{Radial velocity measurements of the \Ion{Na}{i} 8200{\AA}
      absorption doublet from the M star using a Gaussian fit from the 2013
      observations. The red line in the top panel is the centre-of-mass radial
      velocity of the M star determined from our Roche tomography analysis
      (see Section~\ref{sec:roche_tom}). Clear deviations from this fit are
      seen in the residuals in the lower panel as a result of the large number
      of star spots. The red line in the lower panel shows the deviations from
      the centre-of-light to the centre-of-mass radial velocity as determined
      from our Roche tomography analysis of the 2013 data, and show a very
      similar behaviour to the Gaussian measurements.} 
  \label{fig:md_rv}
  \end{center}
\end{figure}

We attempted to measure the radial velocity semi-amplitude of the M star using
the same technique outlined in the last section (i.e. a Gaussian fit) and
fitting the \Ion{Na}{i} 8200{\AA} absorption doublet. Figure~\ref{fig:md_rv}
shows the result of this fit. At first glance it seems that the semi-amplitude
is tightly constrained by the data (the average uncertainty on an individual
velocity measurement is $\sim$1\kms). However, as the residuals show, the
deviation from a pure sinusoid are quite pronounced and there are clearly
large systematic trends affecting the velocity measurements. These deviations
are seen in the data from both 2013 and 2014 as well as for other atomic
features, such as the \Ion{K}{i} 7665{\AA} and 7699{\AA} lines. Therefore, we
do not consider these measurements a true representation of the centre-of-mass
velocity of the M star.

The most likely cause of this deviation is the presence of starspots on the
surface of the M star, although gravity darkening and irradiation may also
contribute to this. Therefore, in order to determine a more reliable value for
the radial velocity semi-amplitude for the M star, as well as a better
understanding of the starspot distribution and evolution, we used our UVES
spectra to perform Roche tomography of the M star, which we detail in the
following section.

\section{Roche tomography} \label{sec:roche_tom}

Roche tomography is a technique analogous to Doppler imaging
\citep[e.g.][]{vogt83}, and has been successfully used to image surface
features on the secondary stars in interacting binaries. The technique assumes
the secondary star rotates synchronously in a circular orbit around the
centre of mass, and uses phase-resolved spectral line profiles to map the
line-intensity distribution across the star's surface in real space. The
method has been extensively applied to the secondary stars in CVs over the
last 20 years
\citep{rutten94,rutten96,watson03,schwope04,watson06,watson07,hill14}, and
artefacts arising from systematic errors are well characterised and
understood. For a detailed description of the methodology and axioms of Roche
tomography, we refer the reader to the references above and the technical
reviews by \citet{watson01} and \citet{dhillon01}.  

\subsection{Least squares deconvolution} \label{sec:lsd}

Spot features appear in absorption line profiles as an emission bump (actually
a lack of absorption), and are typically a few per cent of the line
depth. Thus, very high signal-to-noise ratio (SNR) data are required for
surface imaging, a feat not directly achievable for a single absorption line
with QS\,Vir due to its faintness and the requirements for short exposures to
avoid orbital smearing. To greatly improve our SNR we employ least squares
deconvolution (LSD), a technique that stacks the thousands of stellar
absorption lines observable in a spectrum to produce a single `mean'
profile. First applied by \citet{donati97}, the technique has been widely used
since \citep[e.g.][]{barnes04,shahbaz07}. For a more detailed description of
LSD, we refer the reader to the above references as well as the review by
\citet{collier01}. 

To generate the LSD line profiles the continuum must be flattened. However,
the contribution of the M star to the total system luminosity may vary
due to variations in accretion luminosity (see Section~\ref{sec:evo}), and the
rapid flaring on the M star on timescales of several minutes
\citep{odonoghue03}. Both of these mechanisms may change the continuum slope,
and so a master continuum fit to the data is not appropriate. Furthermore,
normalization of the continuum would result in the photospheric absorption
lines from the secondary varying in relative strength between exposures. This
forces us to subtract the continuum from each spectrum, which was achieved by
fitting a spline to the data.

To produce the LSD profiles we generated a line list appropriate for a M3.5V
star ($\mathrm{T_{eff}} = 3100 \mathrm{K}$ and $\mathrm{log}g = 4.88$) from
the Vienna Atomic Line Database (VALD3; \citealt{kupka00}), adopting a
detection limit of 0.1. The normalized line depths were scaled by a fit to the
continuum of a M3.5V template star so each line's relative depth was correct
for use with the continuum subtracted spectra. As the transitions in molecular
features (such as the prominent TiO bands) are difficult to model, the line
lists for these features are of poor quality, and so we were unable to include
spectral regions containing these features in LSD. In addition, we excluded
emission lines and tellurics from the LSD process. Thus, a total of 46
absorption lines were used, lying in the spectral ranges $8400-8470${\AA} and
$8585-8935$\AA. As QS\,Vir is an eclipsing binary, there are significant
features in the line profiles arising between phases 0.4--0.6 (superior
conjunction) which do not originate on the secondary. As inclusion of these
phases would cause artefacts to appear on the reconstructed maps (e.g. the
entire inner hemisphere would appear to be irradiated), they were excluded
from the fitting process. The LSD profiles were then binned on a uniform
velocity scale of $7.4${\kms} to match the velocity smearing due to the
orbital motion.

The LSD profiles exhibited a clear continuum slope that was largely removed by
subtracting a second-order polynomial fit to the continuum, preserving the
profiles' features and shape. However, these corrected profiles still
exhibited a non-flat continuum which lay significantly below zero. To account
for this, a constant offset was added to each set of LSD profiles, the
specific value of which was non-trivial to determine due to degeneracies in
the fitting process; if a small constant was added, a greater proportion of
the line wings lay below the continuum, inflating the Roche lobe and changing
the optimum masses found in the fit with Roche tomography. If a large constant
was added, more of the line wings lay above the continuum, changing the
optimum system parameters. Hence, an optimum value of constant to add to the
LSD profiles could not be found computationally. Instead, the most appropriate
offset was found by visual inspection of the fits to the data. This was done
independently for each data set. These offset and continuum-flattened LSD
profiles were then adopted for the rest of the fitting process with Roche
tomography.  

\begin{figure*}
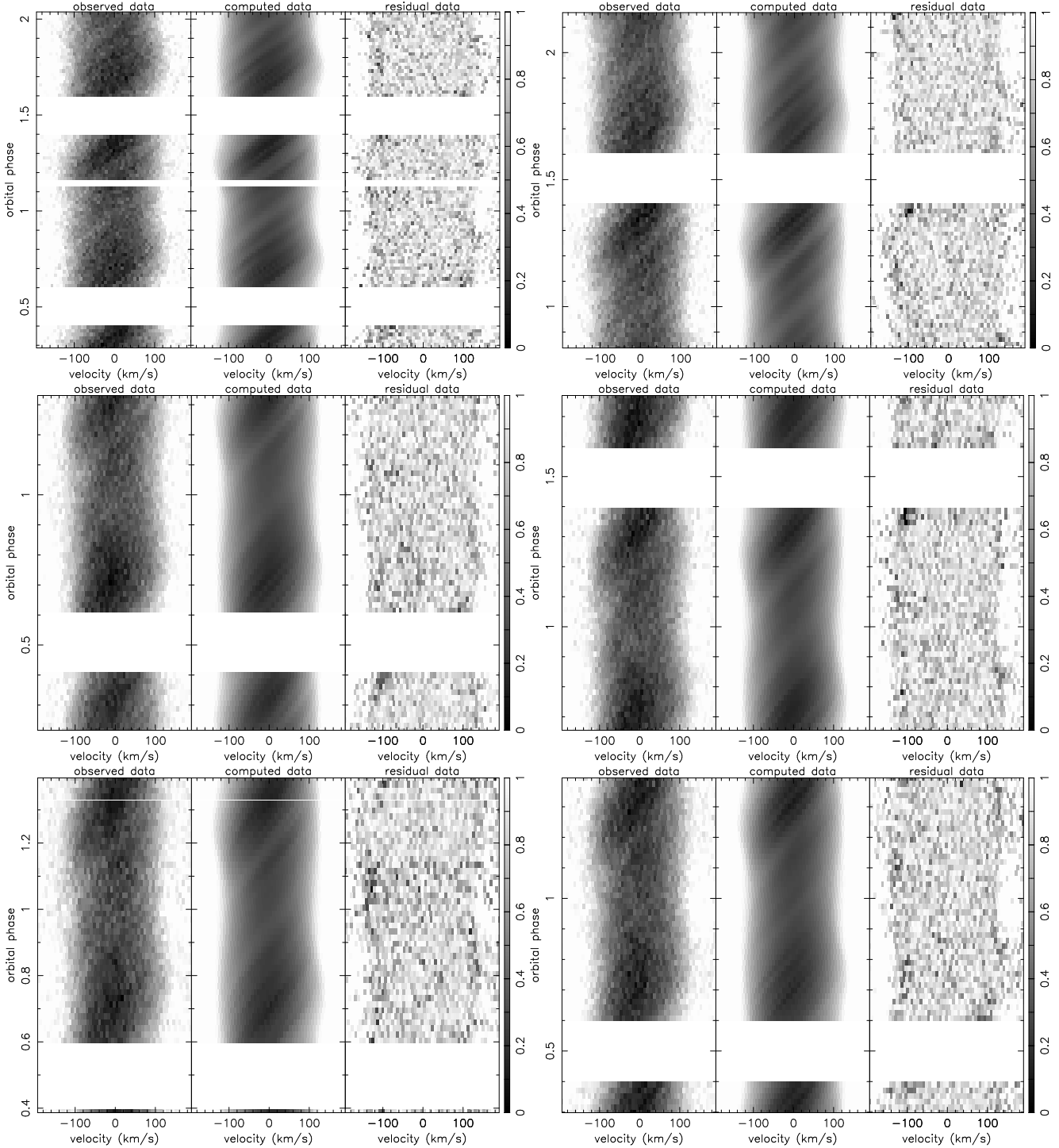

\centering
\includegraphics[width=0.36\textwidth,angle=270]{20130505_trails.eps}
\includegraphics[width=0.36\textwidth,angle=270]{20130506_trails.eps}
\includegraphics[width=0.36\textwidth,angle=270]{20140424_trails.eps}
\includegraphics[width=0.36\textwidth,angle=270]{20140425_trails.eps}
\includegraphics[width=0.36\textwidth,angle=270]{20140430_trails.eps}
\includegraphics[width=0.36\textwidth,angle=270]{20140531_trails.eps}
\caption{Trailed LSD profiles of QS\,Vir (2013-05-04 top left, 2013-05-05 top
  right, 2014-04-23 middle left, 2014-04-24 middle right, 2014-04-31 bottom
  left, 2014-05-31 bottom right). From left to right, panels show the observed
  LSD data, computed data from the Roche tomography reconstruction and the
  residuals (increased by a factor of 4). Starspot features in these panels
  appear bright. A grey-scale wedge is also shown, where a value of 0
  corresponds to the maximum line depth in the reconstructed profiles. The
  orbital motion has been removed assuming the binary parameters found in
  Section~\ref{sec:systempars}, which allows the individual starspot tracks
  across the profiles and the variation in $V_{rot}\sin{i}$ to be more clearly
  observed.} 
\label{fig:trails}
\end{figure*}

 In general, Roche tomography cannot normally be performed on data which
  has not been corrected for slit losses, due to the secondary star's variable
  light contribution. As the acquired data of QS\,Vir was largely photometric,
  we were able to correct for slit losses using an optimal-subtraction
  technique. Here, each LSD profile was scaled and subtracted from the fit
  made with Roche tomography, where the optimal scaling factor was that which
  minimized the residuals. The newly-scaled LSD profiles were then re-fit with
  Roche tomography, and the above process repeated until the scaling factors
  no longer varied. The resulting scaled profiles were visually inspected and
  found to be consistent. The final LSD profiles with computed fits and
residuals are shown in Figure~\ref{fig:trails}. 

\subsection{System parameters} \label{sec:systempars}

Roche tomography constrains the binary parameters by reconstructing intensity
maps of the stellar surface for many combinations of $M_{1}$, $M_{2}$,
inclination ($i$), systemic velocity ($\gamma$) and Roche-lobe filling factor
(RLFF). Each map is fit to the same $\chi^{2}$. However, for a given value of
$\chi^{2}$, many maps may fit the data equally well, and so the
maximum-entropy regularisation statistic is employed for selecting the
reconstructed map with least informational content required to fit the data
(the most positive entropy value). Thus for any given set of parameters, the
reconstructed map has an associated entropy. The optimal parameters are
determined by selecting those that produce the map of maximum entropy across
all possible parameter combinations. Adoption of incorrect system parameters
results in the reconstruction of spurious artefacts in the final map (see
\citealt{watson01} for details). Artefacts are well characterised and always
increase the amount of structure mapped in the final image, which in turn
leads to a decrease in the entropy regularisation statistic. All data sets
were fit independently using this method. 

It was assumed the rotation of the M dwarf is tidally locked to the orbital
period, and so the period was fixed to 0.1507576116~d
\citep{parsons11}. Additionally, the inclination was fixed to $i =
77.8^{\circ}$, as this was well constrained from light-curve analysis --
considering the SNR of the data being fit, we are unlikely to improve the
accuracy of this using Roche tomography, and indeed an unsuccessful attempt
was made to constrain inclination once optimal system parameters were found. 

\subsubsection{Limb Darkening} \label{sec:limbdarkening}

Roche tomography allows limb darkening to be included in the fitting of line
profiles. In order to calculate the correct limb darkening coefficients, we
determined the effective central wavelength of the data over ranges specified
in Section~\ref{sec:lsd} using,
\begin{equation}
\lambda_{\mathrm{cen}} = \frac{\sum_{i}\nolimits \frac{1}{\sigma_{i}^2}\:d_{i}\:\lambda_{i}}{\sum_{i}\nolimits\:\frac{1}{\sigma_{i}^2}\:d_{i}},
\label{eq:cenwave}
\end{equation}
where $d_{i}$ is the line depth at wavelength $\lambda_{i}$, and $\sigma_{i}$
the error in the data at $\lambda_{i}$. The mean value of $\lambda_{cen}$
across the 6 datasets was used for consistency. We adopted a four-parameter
non-linear limb darkening model (see \citealt{claret00}), given by 
\begin{equation}
\frac{I(\mu)}{I(1)} = 1 - \displaystyle\sum_{k=1}^{4}a_{k}(1-\mu^{\frac{k}{2}}),
\label{eq:limbdarkeninglaw}
\end{equation}
where $\mu = \cos{\gamma}$ ($\gamma$ is the angle between the line of sight
and the emergent flux), and $I(1)$ is the monochromatic specific intensity at
the centre of the stellar disk.  

Using the calculated $\lambda_{cen}$, the limb darkening coefficients were
then determined by linearly interpolating between the tabulated wavelengths
given in \citet{claret00}. We adopted stellar parameters closest to that of a
M3.5V star, which for the PHOENIX model atmosphere were $\log{g} = 5$ and
$T_{\mathrm{eff}} = 3100 \mathrm{K}$. Thus, the adopted coefficients were
$a_{1}=3.302, a_{2}=-5.057, a_{3}=4.612, a_{4}=-1.61$. We note that the
  limb darkening law used is for spherical stars, and so is not ideally suited
  to the distorted main-sequence star in QS\,Vir. Ideally one would need to
  calculate limb darkening coefficients for each tile in the grid based on a
  model atmosphere. However, this is not currently possible within our code,
  and so we adopted uniform limb darkening coefficients. Artefacts arising
  from adopting incorrect coefficients are discussed in \citet{watson01}.

\subsubsection{Systemic velocity} \label{sec:systemicvelocity}

\begin{figure}
\centering
\includegraphics[width=0.45\textwidth]{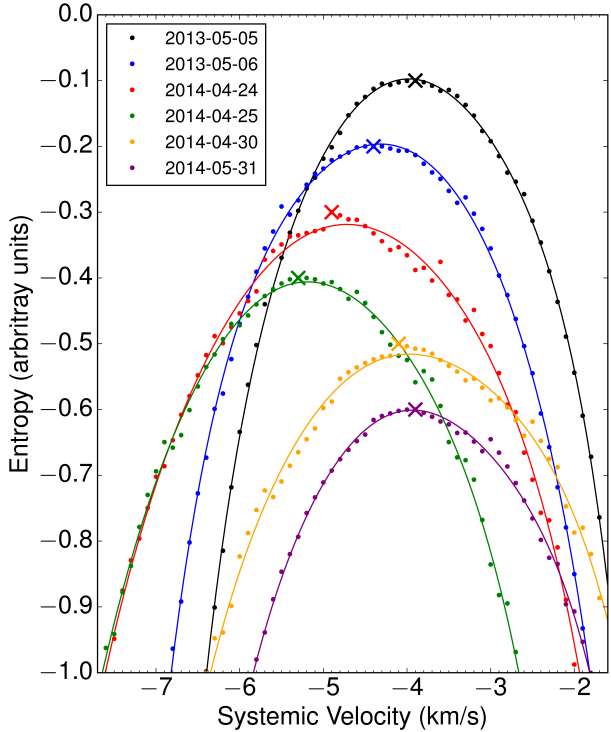}
\caption{Points show the maximum entropy value obtained in each data set, as a
  function of systemic velocity. The optimal masses and RLFF as found in
  Section~\ref{sec:rlffmasses} were adopted for the fits of each data set. The
  points are plotted with a vertical offset of 0.1 between data sets. Crosses
  mark the value of $\gamma$ that yields the map of maximum entropy, and a
  sixth-order polynomial fit is shown as a visual aid. The mean systemic
  velocity across all data is $\gamma_{\mathrm{mean}} =
  -4.4_{-0.9}^{+0.5}$~kms$^{-1}$.}
\label{fig:systemic}
\end{figure}

The entropy statistic is most sensitive to an incorrect systemic velocity
$\gamma$, and so this parameter was narrowed down most
easily. Figure~\ref{fig:systemic} shows the map entropy yielded for each data
set as a function of $\gamma$, where a cross marks the $\gamma$ that gives the
map of maximum entropy. The mean value across all data sets yields
$\gamma_{\mathrm{mean}} = -4.4_{-0.9}^{+0.5}$\kms, where the
uncertainties represent the spread between maximum and minimum for all
measured values. We also determined the radial velocity semi-amplitude of the
M star as $275.8 \pm 2.0$\kms. Note that the radial
  velocity semi-amplitude of the M star is calculated from the resulting
  stellar parameters (masses, inclination etc.) rather than being a direct
  input to the fit. As mentioned in Section~\ref{sec:rvmd} we
consider these values more reliable than those from Gaussian fitting, due to
the surface inhomogeneities biasing those results and indeed once these
  surface features are taken into account, the measured velocities match
the reconstructed model from Roche tomography (see the lower panel of
Figure~\ref{fig:md_rv}). 

\subsubsection{Roche lobe filling factor and masses} \label{sec:rlffmasses}

\begin{figure}
\centering
\includegraphics[width=0.45\textwidth]{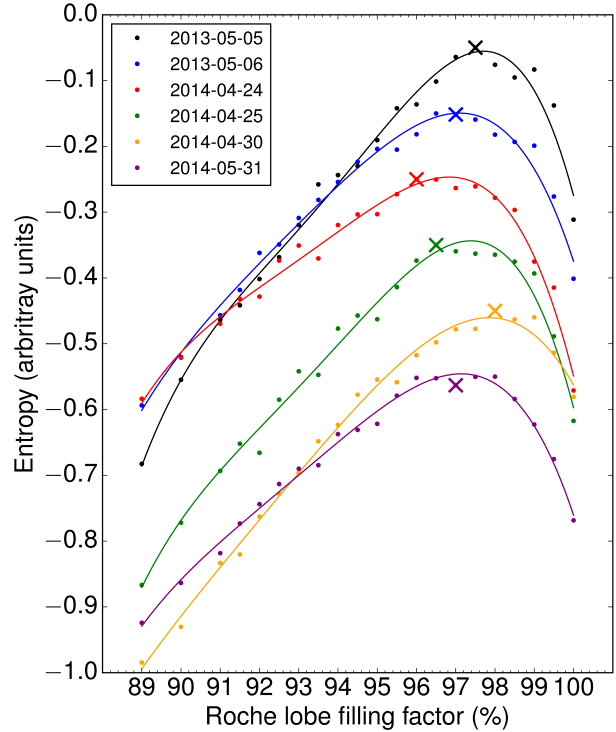}
\caption{Points show the map entropy obtained in each data set as a function
  of RLFF, where optimal masses were used for each RLFF. The optimal systemic
  velocities found in Section~\ref{sec:systemicvelocity} were adopted for the
  map reconstructions. The points are plotted with a vertical offset of 0.1
  between data sets. Crosses mark the maps of maximum entropy, yielding the
  optimal masses at the optimal RLFF. Fourth-order polynomial fits are shown
  as a visual aid. The peaks of these fits were \emph{not} used to determine
  the optimal RLFF, as the high-order nature of the polynomial meant the fits
  were often skewed. The mean RLFF across all data is $97\pm1$~per~cent, where
  the uncertainty represents the spread between maximum and minimum for all
  measured values.}
\label{fig:ffmass}
\end{figure}

Readers should note that, unless otherwise stated, the value of the RLFF is
given as the ratio of the volume averaged radius of the M star to its Roche
lobe radius, as opposed to the linear RLFF which is given by the distance from
the centre of mass of the M star to its surface in the direction of the L1
point divided by the distance to the L1 point. 

Figure~\ref{fig:ffmass} shows the map entropy as a function of RLFF. For a
given RLFF, the optimal masses are determined by reconstructing maps for many
pairs of component masses. The optimal $M_{1}$ and $M_{2}$ are those that
produce the map of maximum entropy (an `entropy landscape', see Figure~8 of
\citealt{hill14}). The values of $\gamma$ determined in
Section~\ref{sec:systemicvelocity} were adopted for the fits, and
reconstructions were performed in steps of 0.5~per~cent for the RLFF, and in
steps of 0.1{\MSUN} for the masses. The points follow a roughly parabolic
trend with some small `jumps' in entropy that are mainly due to the change in
grid geometry, due to varying masses and RLFF, combined with rounding errors
in the code. However, all optimum RLFFs determined from each independent data
set are in close agreement, ranging between 96--98~per~cent, and so we can be
confident that the measured RLFF and masses are robust. The amount of scatter
in the entropy `parabolas' is consistent with the low SNR of the data being
fit -- similar scatter was found when performing Roche tomography with
relatively low-quality data of the CVs AM~Her and QQ~Vul \citep{watson03}. 

There are a number of possible sources of systematic error when fitting data
of poor SNR with Roche tomography. In this instance, the dominant source of
uncertainty in the system parameters is due to the difficulty in determining
the true continuum level of the LSD profiles, as discussed in
Section~\ref{sec:lsd}. If the continuum is too low, the Roche lobe is
artificially inflated to fit the line wings, resulting in larger measured
masses. Furthermore, if the continuum is set too high, the Roche lobe is
artificially under-filled. To determine the impact of choosing an incorrect
continuum level, a range of offsets were added to the LSD profiles, and for
each offset, the LSD profiles were fit and scaled as described in
Section~\ref{sec:lsd}. The optimal masses found for each offset were then
compared and, across all data, the maximum difference between the optimal
primary and secondary masses were 3.8~per~cent and 7.9~per~cent, respectively,
and the spread in RLFF was 1~per~cent. Hence, while only a visual check was
made of the fits to the continuum level of each set of LSD profiles, the
optimal masses and RLFF found for each data set agree within these
uncertainties. 

The mean RLFF is $97\pm1$~per~cent, with the mean of the masses yielding
$M_\mathrm{WD}=0.76_{-0.01}^{+0.02}${\MSUN} and
$M_\mathrm{dM}=0.358_{-0.018}^{+0.022}$\MSUN, respectively, where all
uncertainties represent the spread between the maximum and minimum
values. Hence, the mean mass-ratio was found to be $q = 0.47\pm0.04$. In
Section~\ref{sec:lc_model} we will combine the spectroscopic constraints with
our light curve data to determine more precise values for these parameters,
but quote these Roche tomography values here for completeness. Using
the mean system parameters with our model grid, we calculate $v\sin{i}$ to
range between a minimum of 117.8{\kms} at phase 0 and 0.5, to a maximum
of 133.5{\kms} at phase 0.25 and 0.75. Our measured masses and RLFF agree
well across all data sets, and so for consistency in comparing maps, we
adopted the mean values for the final map reconstructions (see
Figure~\ref{fig:mapallpbin}).  

\begin{figure}
\centering
\includegraphics[width=\columnwidth]{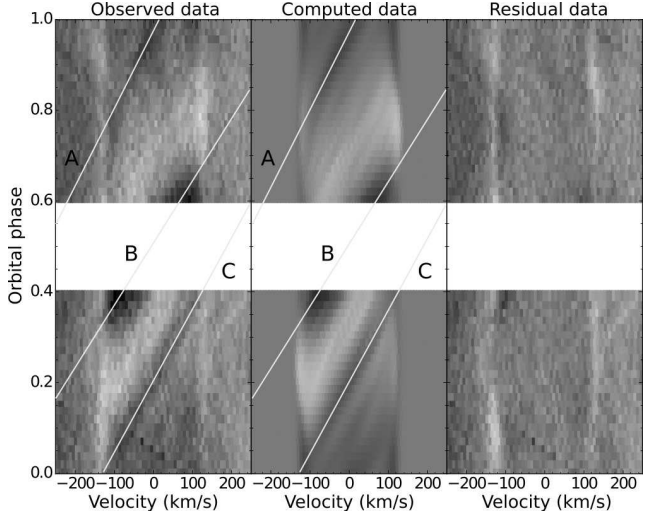}
\caption{Trailed LSD profiles of QS\,Vir using phase-binned data from all
    data sets. A mean of the computed LSD profiles has been subtracted from
    both the observed and computed LSD profiles. Starspot features in these
    panels appear dark. The orbital motion has been removed assuming the
    binary parameters found in Section 4.2.}
\label{fig:trailsallpbin}
\end{figure}

\subsection{Surface maps} \label{sec:surfacemaps}

\begin{figure}
\centering
\includegraphics[width=\columnwidth]{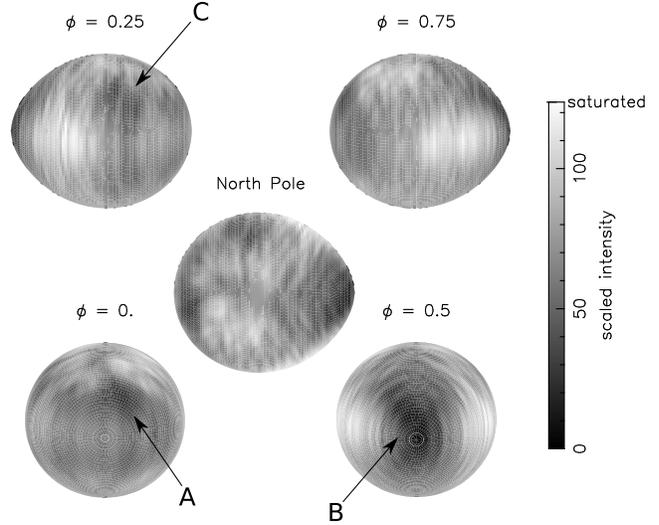}
\caption{Roche tomogram of QS\,Vir using phase-binned data from all data
  sets. Dark grey scales indicate regions of reduced absorption line strength
  that is due to either the presence of starspots or the impact of
  irradiation. The contrast in these Roche tomograms has been enhanced such
  that the darkest and lightest grey-scales correspond to the lowest and
  highest intensities in the reconstructed map. The orbital phase is indicated
  above each panel, and Roche tomograms are shown without limb darkening for
  clarity.} 
\label{fig:mapallpbin}
\end{figure}

The mean system parameters found in Section~\ref{sec:systempars} were used to
construct Roche tomograms of all data sets. The corresponding trails of LSD
profiles are displayed in 
Figure~\ref{fig:trails}. Evident in both the trails and the maps are numerous
spot features, with three large-scale features common to all trails and maps
(marked A, B and C in
Figures~\ref{fig:trailsallpbin}~and~\ref{fig:mapallpbin}). The technique
  of Roche tomography is shown to be remarkably robust in the reconstruction
  of these features  -- maps reconstructed independently
with data taken on sequential nights show very similar features. The
morphologies of these three features remain fairly unchanged over the 1 year
period of observations, with no apparent movement across the stellar surface,
indicating that these are indeed long-lived starspots. Furthermore,
smaller-scale structure is apparent in both the data and fits from the 2013
data, however the same level of detail is not as readily reconstructed using
other data sets, most likely due to the fewer number of LSD profiles being
fit. 

There is a distinct lack of a large high-latitude spot. Such a feature is
commonly seen on Doppler imaging studies of rapidly rotating single stars
\citep[e.g.][]{donati99,hussain07}, and in CVs such as BV~Cen and AE~Aqr 
\citep{watson07,watson06}. Given the presence of such a feature in
these other rapidly rotating stars, it is surprising to find that a large
high-latitude spot does not exist here. Pertinently, a study by
\citet{morin08} finds that large-scale, mainly axisymmetric poloidal fields
are fairly common in fully convective M dwarfs, observing polar region spots
on three out of the six stars in their sample. However, the authors also found
that the partly convective star AD~Leo hosted a similar magnetic field, but
with significantly lower magnetic flux, indicating the generation of
large-scale magnetic fields is more efficient in fully convective stars. As
the measured mass of the M dwarf in QS\,Vir suggests it has a partly radiative
core, the lower efficiency of magnetic field generation may mean any polar
region spots are significantly less obvious. 

Alternatively, the lack of a large high-latitude feature may simply be due
observational effects -- given the inclination of $i = 77.8\degr$ in QS\,Vir,
high latitudes appear severely foreshortened, and are heavily
limb-darkened. These effects, combined with the poor SNR of the data, may mean
high-latitude features are simply lost in the noise. This was assumed to be
the case by \citet{barnes01} who found a similar lack of polar spot features
in their study of HK~Aqr ($i = 70-90^{\circ}$), with simulations showing that
reconstructions of high-latitude features strongly depend on the SNR of the
data. 

Other prominent features on the maps are the dark regions around the L$_{1}$
point, marked `B' in
Figures~\ref{fig:trailsallpbin}~and~\ref{fig:mapallpbin}. In previous maps of
CVs (e.g. \citealt{watson03}) features on this part of the star were
attributed to irradiation effects from the white dwarf, where ionisation of the
photosphere causes a decrease in photometric flux. However, the features
reconstructed here are patchy in morphology, rather than smoothly varying,
which suggests that irradiation is not the main contributing
factor of the features in this region. This was tested by simulating a
realistic irradiation pattern, and fitting the data using the same phases as
that of the phase-binned LSD profiles shown in
Figure~\ref{fig:trailsallpbin}. The reconstructed irradiation pattern was
found to be smoothly varying, and was essentially uniform around the front of
the star. Likewise, the effect of gravity darkening on map reconstruction was
simulated using a gravity darkening coefficient of $\beta \sim 0.1$, as found
by \citet{djurasevic03,djurasevic06} for late-type M stars. The
simulated map showed the effects of gravity darkening to be smoothly varying,
with minimal contribution to the mapped features. In addition, the dominant
features in the line profiles between phases 0.4--0.6 (as noted in
Section~\ref{sec:lsd}) were thought to originate off the stellar surface, and
so these phases were excluded from the fit. Hence it is unlikely that
circumstellar material is the source of these patches around the L$_{1}$
point.

We conclude that there is significant spot coverage on the inner
hemisphere (white dwarf facing side of the M star) observed on all maps. This
spot distribution -- where there is a higher density of spots on the
hemisphere facing the companion star -- is similar to that observed in many
other systems, such as the pre-CV V471 Tau \citep{hussain06}, the eclipsing
binary ER~Vul \citep{xiang15}, and the close binary $\sigma^{2} \text{CrB}$
\citep{strassmeier03}. Furthermore, \citet{kriskovics13} suggested that the
hot-spots mapped on the companion-facing hemisphere in V824~Ara may indicate
the strong interaction between the magnetic fields of the component stars, as
is commonly observed in close RS CVn-type binaries. This phenomenon was
discussed in \citet{hill14}, where spot `chains' from the pole to L$_{1}$
point (on multiple CVs) suggested a mechanism that forces magnetic flux tubes
to preferentially emerge at these locations. Indeed, \citet{holzwarth03}
propose that tidal forces may cause spots to form at preferred longitudes, and
\citet{moss02} suggest that this phenomena may also be due to the tidal
enhancement of the dynamo action itself.

Other features of interest include a prominent starspot labelled `A' in
Figure~\ref{fig:mapallpbin}. It stretches $\sim5$--25$^{\circ}$ in longitude, and
$\sim15$--35$^{\circ}$ in latitude (where the back of the star is at $0^{\circ}$
longitude, with increasing longitude in the direction of the leading
hemisphere). Finally, label `C' in Figure~\ref{fig:mapallpbin} points to a
group of spots covering 275--300$^{\circ}$ longitude and spanning 20--55$^{\circ}$
latitude. Given that the phase-binned data has a longitudinal resolution of
$\sim6^{\circ}$, we can reliably separate the larger spots in this group -- one
spot extends 275--285$^{\circ}$ longitudinally and 35--55$^{\circ}$ in
latitude, and another spans 293--300$^{\circ}$ in longitude and
15--30$^{\circ}$ latitudinally. As both `A' and `C' can be seen in all data
trails taken over a $\sim1$~yr period, they are most likely groups of
long-lived starspots.

 The prominent spotted regions labelled `A' and `B' (located at phases
  $\phi \sim 0.05$ and $\phi \sim 0.5$, respectively) are similar to those
  found by \citet{ribeiro10} in their maps of QS\,Vir made using photometry
  taken in 1993 and 2002. The authors find two cool regions at $\phi \sim 0.4$
  and $\phi \sim 0.9$ that show little shift in longitude between 1993 and
  2002. The authors' photometric data does not allow for a tight constraint on
  latitude, however, if these spotted regions are the same as those imaged here
  (namely features `A' and `B'), they may be very long lived active regions
  that have remained at the same longitudes for over 20 years. 

\subsubsection{Spot coverage as a function of longitude and latitude}

To make a more quantitive estimate of spot parameters on QS\,Vir, we have
examined the pixel intensity distribution in the Roche tomogram. By visual
inspection, all reconstructed maps appeared to have a very similar
distribution of spots, with small-scale variation due to differences in SNR
and number of spectra between data sets. For an improved estimate of spot
coverage, we combined all data sets by phase-binning the LSD profiles over 60
evenly-spaced phase bins (each containing $\sim7$ LSD profiles). We only
present analysis of the map that was reconstructed using this binned data (see
Figure~\ref{fig:mapallpbin}). 

The high inclination of the system causes a `mirroring' effect, as described
in section 3.11 of \citet{watson01}. As radial velocities cannot constrain
whether a feature is located in the Northern or Southern hemisphere, features
are mirrored about the equator. This means features become less apparent due
to the fact that they are first latitudinally smeared, and then mirrored,
causing the streaks seen in the maps presented here. To mitigate this effect,
we discarded all pixels in the Southern hemisphere for the remainder of this
analysis.

Each pixel in our Roche tomograms was given a spot filling-factor between 0
(immaculate) to 1 (totally spotted). The value for any given pixel depended on
its intensity, and was scaled linearly between our predefined immaculate and
spotted photosphere intensities. It is unlikely that the highest intensities
represent the immaculate photosphere, as the growth of bright pixels in maps
that are not `threshholded' is a known artefact of Doppler imaging techniques
\citep[e.g.][]{hatzes92}. Instead, we have defined an intensity of 92 and
above to represent the immaculate photosphere. Furthermore, we have
defined a 100~per~cent spotted pixel by taking the minimum pixel intensity in
the group of spots marked `B' in Figure~\ref{fig:mapallpbin}, avoiding the
very front pixels on the model star as these may be additionally affected by
gravity darkening or irradiation. 

Based on our classification of a spot, we estimate that 35~per~cent of the
northern hemisphere of QS\,Vir is spotted. This is likely to be a lower limit
due to the presence of unresolved spots -- \citet{oneal98} estimated up to
50~per~cent fractional spot coverage in TiO studies of rapidly-rotating G and
K type stars, and \citet{senavci15} find a factor~$\sim10$ difference between
spot filling factors when comparing TiO analysis to Doppler
imaging. Furthermore, \citet{hessman00} predict a mean spot coverage of
$\sim50$~per~cent on the CV, AM~Her, using long-term light curve analysis.

\begin{figure}
\centering
\includegraphics[width=0.5\textwidth]{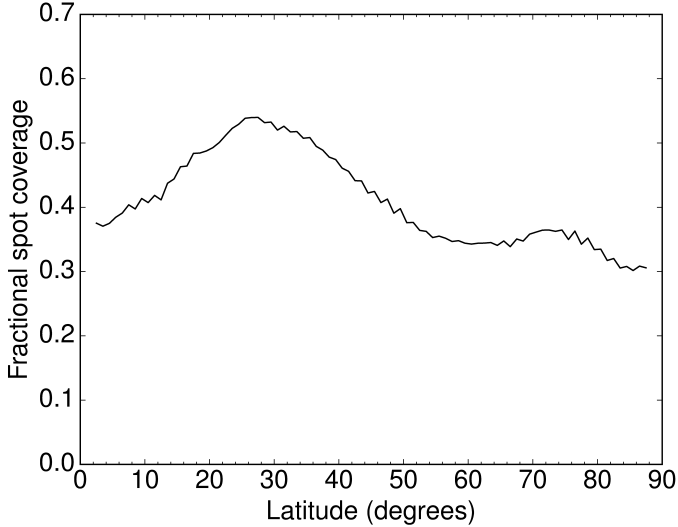}
\caption{Fractional spot coverage as a function of latitude, normalised by the
  surface area of that latitude, as calculated for the reconstructed map using
  phase-binned data. The geometry the grid in our Roche tomogram does not
  allow for strips of constant latitude, and so we are required to interpolate
  between grid elements. We have done this by taking a moving-mean with a
  $5^{\circ}$ window in steps of $1^{\circ}$.}
\label{fig:allpbinlat}
\end{figure}

When plotted as a function of latitude (see Figure~\ref{fig:allpbinlat}) we
see a clear increase in fractional spot coverage at low to mid latitudes
(10--50$^{\circ}$) of around 0.2, as compared to the spot coverage at higher
latitudes (50--90$^{\circ}$). This large spread in latitude may be indicative of
a distributed dynamo, with many unresolved small-scale features spread
across the stellar surface. A more likely explanation, however, is that
  the spread is due to features becoming smeared in latitude due to both poor
  SNR, and from limitations in the technique (as discussed above). Also
apparent is a small increase in spot coverage at high latitudes (centred
$\sim74^{\circ}$). As discussed above, the spot contrast at higher latitudes
is diminished in this system due to is high inclination, and so the measured
spot coverage is likely underestimated at these latitudes. Similar
distributions of spots were found on the M~dwarfs HK~Aqr and RE~1816+541 by
\citet{barnes01}, with the majority of features equally distributed over
20--80$^{\circ}$~latitude.  

\begin{figure}
\centering
\includegraphics[width=0.5\textwidth]{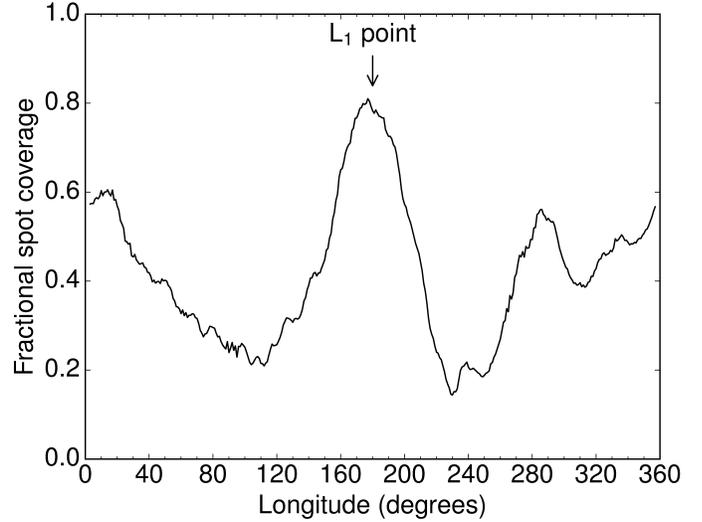}
\caption{Fractional spot coverage as a function of longitude, as calculated
  for the reconstructed map using phase-binned data. Points are calculated by
  taking a moving-mean with a $6^{\circ}$ window (the longitudinal resolution
  of the map), in steps of $1^{\circ}$.}
\label{fig:allpbinlong}
\end{figure}

When plotted as a function of longitude in Figure~\ref{fig:allpbinlong}, we
find a large increase in fractional spot coverage around $15^{\circ}, 180^{\circ}$,
and~$285^{\circ}$ longitude, corresponding to the features marked A, B and C,
respectively, as shown in Figure~\ref{fig:mapallpbin}. As previously
discussed, the large increase around $180^{\circ}$ is similar to that found on
CVs (e.g. AE Aqr, \citealt{hill14}) and on pre-CVs (e.g. V471 Tau,
\citealt{hussain06}). 

\subsubsection{Magnetic activity and star spot lifetime}

The features marked A, B and C in
Figures~\ref{fig:trailsallpbin}~and~\ref{fig:mapallpbin} are clearly observed
in all data sets from May 2013 to April 2014, and appear to remain in fixed
positions over both short and long timespans. This suggests a very low level
of surface shear, with no perceivable shift over the $\sim 1$~yr between first
and last observations. As the M star in QS\,Vir is very close to
becoming fully convective, the lack of any detectable differential rotation
concurs with the conclusions of \citet{barnes05}, who found that differential
rotation appears to vanish with increasing convection depth. This trend was
also found in studies of rapidly-rotating M~dwarfs by \citet{morin08}, where
several stars in their sample exhibited fairly similar magnetic topologies
over the $\sim 1$~yr period between observations. In other work,
\citet{morin10} found both GJ\,51 and WX~UMa to exhibit strong, large-scale
axisymmetric-poloidal and nearly dipolar fields, with very little temporal
variations over the course of several years.

We see no evidence of a magnetic activity cycle in our
observations. \citet{odonoghue03} suggested that the $\pm12$ s spread on
ephemerides over their 10 years of eclipse timings may be due to such a
cycle. This appears unlikely, since their Figure~3~(b) shows a
$\sim0.12$~per~cent period change in $\sim 600$~d -- One would expect spot
migration to occur over the course of a magnetic activity cycle, as seen on
the Sun, and as we find spot features are stable over a $\sim1$ yr period,
such a cycle cannot be responsible for this rapid change in period.  

These features may well be related to the large prominences that pass in front
of both stars. Since the prominences appear to be stable on a timescale of
years, they must still be anchored to the surface of the M star by its
magnetic field. Starspots are likely to form near these anchor regions, as is
seen in solar coronal loops and prominences. 

\section{Modelling the eclipse light curve} \label{sec:lc_model}

Eclipsing PCEBs have been used to measure precise masses and radii of both
the white dwarf and main-sequence components in a number of systems
\citep[e.g.][]{parsons10nn,parsons12hipr,parsons12cool} offering some of the
most precise and model-independent measurements to date for either type of
star. The small size of the white dwarf leads to very sharp eclipse features,
with typical ingress and egress times of around a minute or less meaning that
high-speed photometry is required to properly resolve the eclipse
features. With typical exposure times of $<3$ sec, our ULTRACAM data fully
resolve the eclipse of the white dwarf and thus can be used to place stringent
constraints on the physical parameters of both stars.

\begin{figure*}
  \begin{center}
    \includegraphics[width=\textwidth]{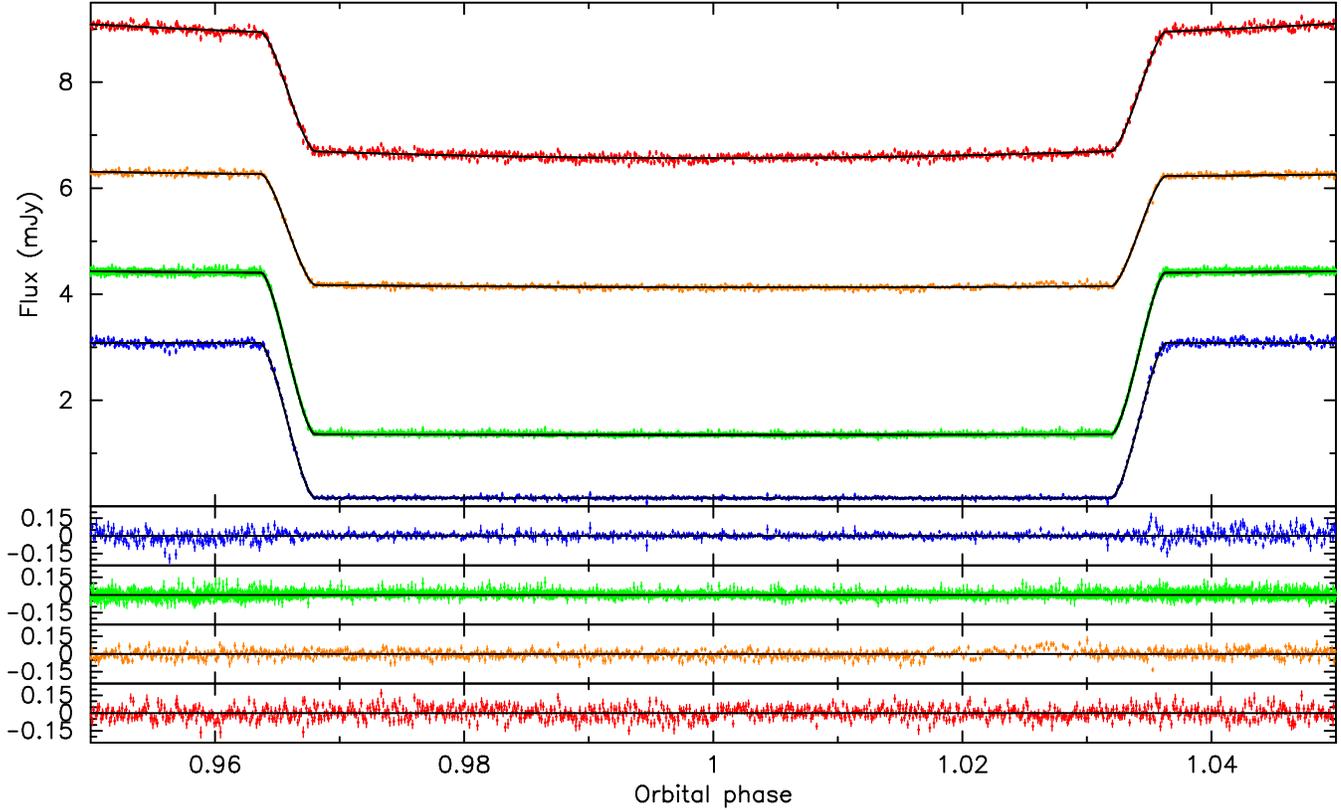}
    \caption{Flux calibrated ULTRACAM $u'g'r'$ and $i'$ band light curves of
      the eclipse of the white dwarf in QS\,Vir with model fits overplotted
      (black lines). We have offset the $i'$ band light curve by -13mJy to
      show more detail. The bottom four panels show the residuals of the fit in
      the $u'$ band (top panel, blue points), $g'$ band (second panel, green
      points), $r'$ band (third panel, orange points) and the $i'$ band
      (bottom panel, red points).}
  \label{fig:lc_fit}
  \end{center}
\end{figure*}

We model the light curve data using a code written for the general case
of binaries containing white dwarfs (see \citealt{copperwheat10} for a
detailed description). The program subdivides each star into small elements
with a geometry fixed by its radius as measured along the direction of centres
towards the other star with allowances for Roche geometry distortion. We
fitted the ULTRACAM $u'g'r'$ and $i'$ band light curves around the eclipse 
of the white dwarf. We exclude data taken at other orbital phases since the
effects of starspots on the surface of the main-sequence star strongly
influences the constraints that these data provide. However, the effects of
starspots on the white dwarf eclipse itself are minor, since it is such a
rapid process. Nevertheless, starspots often lead to linear slopes across the
eclipse, that vary over the timescale of our ULTRACAM observations. Therefore,
each eclipse observation was initially fitted with a basic model using the
parameters from \citet{parsons10time} and including a linear slope. This slope
was then removed from each light curve individually before combining
them. Additionally, we shifted each eclipse in phase to remove the effects of
the deviations from linearity in the eclipse arrival times.

The basic geometric parameters required to define the model are the mass
ratio, $q = M_\mathrm{sec}/M_\mathrm{WD}$, the inclination, $i$, the scaled
radii of both stars, $R_\mathrm{WD}/a$ and $R_\mathrm{sec}/a$ and the two
temperatures, $T_\mathrm{eff,WD}$ and $T_\mathrm{eff,sec}$. In addition to
these, the model also requires the time of mid eclipse, $T_{0}$, the period,
$P$ and the limb darkening parameters for both stars.

Since we used phase-folded data we kept the period fixed as 1, but allowed
$T_{0}$ to vary. The temperature of the white dwarf was fixed at
$T_\mathrm{eff}=$14,220\,K. We use the same setup for limb darkening as
detailed in Equation~\ref{eq:limbdarkeninglaw} for both stars. For the M star,
we use the limb darkening coefficients for a $T_\mathrm{eff}=$3,100\,K,
$\log{g}=5$ main sequence star in the appropriate filter \citep{claret00}. We
use the coefficients for a $T_\mathrm{eff}=$14,000\,K, $\log{g}=8.25$ white
dwarf from \citet{gianninas13}. All these limb darkening coefficients were
kept fixed during the fitting process.

The profile of the eclipse of the white dwarf does not contain enough
information to simultaneously constrain the inclination and the scaled radii
of both stars, only how they relate to each other. Therefore, an additional
constraint is required to break this degeneracy. Ideally we would use the
secondary eclipse (the transit of the white dwarf in front of the M star, see
\citealt{parsons10nn} for example), but for QS\,Vir this is too shallow to
detect in our ULTRACAM data. An alternative method would be to use the
amplitude of the ellipsoidal modulation, which is related to the mass ratio
(which we have already spectroscopically constrained) and the scaled radius of
the M star \citep[e.g.][]{parsons12cool}. However, the effects of starspots
make this approach unreliable, even with the Roche tomography
results. Moreover, this limits our fits to the longer wavelength bands where
the M star dominates and hence where the white dwarf eclipse is shallower and
offers poorer constraints on its size. Therefore, this approach was not taken,
although we did check that our final model results predicted an ellipsoidal
modulation amplitude consistent with the ULTRACAM data. Another method used to
break the degeneracy between the inclination and scaled radii, and that we use
here, is to use our measurement of the rotational broadening of the M star
($V_\mathrm{rot}\sin{i}$). For a synchronously rotating star this is given by
\begin{equation}
V_\mathrm{rot}\sin{i} = K_\mathrm{sec}(1+q)\frac{R_\mathrm{sec}}{a},
\end{equation}
where the radial velocity semi-amplitude of the M star ($K_\mathrm{sec}$) and
the mass ratio ($q$) have both been spectroscopically constrained. Hence,
combining this with the fit to the white dwarf eclipse allows us to determine
the inclination, radii of both stars and, via Kepler's laws, the orbital
separation and masses of both stars. However, as previously noted in
section~\ref{sec:rlffmasses}, the measured rotational broadening varies as a
function of orbital phase, as the M star is not spherical and hence presents a
different radius at different orbital phases. Therefore, when fitting the
eclipse light curve we calculated the rotational broadening of the M star at
the quadrature phases (when it is at its maximum) and compared this to the
maximum value determined from our Roche tomography analysis.

We used the Markov Chain Monte Carlo (MCMC) method to determine the
distributions of our model parameters \citep{press07}. The MCMC method
involves making random jumps in the model parameters, with new models being
accepted or rejected according to their probability computed as a Bayesian
posterior probability. In this instance this probability is driven by a
combination of the $\chi^2$ and the prior probability from our constraints
from the spectroscopy and Roche tomography. For each band an initial MCMC
chain was used to determine the approximate best parameter values and
covariances. These were then used as the starting values for longer chains
which were used to determine the final model values and their
uncertainties. Several chains were run simultaneously to ensure that they
converged on the same values.

This approach results in very few systematic uncertainties in the derived
  parameters, since the vast majority of them are directly determined by the
  data itself. The temperatures of the two stars are effectively scale factors
  and so have no effect. The shape of the eclipse of the white dwarf is not
  altered by the adopted limb darkening coefficients for the M star, these
  are mostly important for the transit of the white dwarf across the face of
  the M dwarf as well as any ellipsoidal modulation, neither of which we
  consider in our fit. However, the choice of limb darkening coefficients for
  the white dwarf does have some effect on the final parameters, specifically
  the final radius measurement for the white dwarf (and to a lesser extent the
  M dwarf). To determine how large this effect is we re-fitted the eclipse
  light curves using limb darkening coefficients for a white dwarf with
  $T_\mathrm{eff}=13,500$\,K and $T_\mathrm{eff}=14,500$\,K and $\log{g}=8.00$
  and $\log{g}=8.50$, taken from \citet{gianninas13}, these bracket the
  coefficients that we used in our final fit. We found that the differences in
  the final parameters were smaller than their statistical uncertainties,
  implying that the white dwarf's limb darkening coefficients have a very
  minor effect on the fit.

  The only other source of systematic uncertainty in the light curve fit arises
from the Roche tomography analysis. As previously mentioned, we used the
rotational broadening and M dwarf radial velocity semi-amplitude measurements
from the Roche tomography analysis to help break the degeneracies in the light
curve fit. However, as Figures~\ref{fig:systemic} and \ref{fig:ffmass} show,
the values vary slightly from night to night. This was taken into account by
increasing the uncertainties on these two values to cover the spread of values
seen on different nights. Indeed, this is the largest source of uncertainty in
the fit and limits how precisely we could measure the stellar masses.

\begin{table}
 \centering
  \caption{Stellar and binary parameters of QS\,Vir. The orbital period and T0
  values were taken from \citet{parsons10time}, the distance and white dwarf
  temperature values were taken from \citet{odonoghue03}.}
  \label{tab:params}
  \begin{tabular}{@{}lc@{}}
  \hline
  Parameter                         & Value  \\
  \hline
  RA                                & 13:49:51.95 \\
  Dec                               & -13:13:37.5 \\
  Orbital period                    & 0.150\,757\,463(4)\\
  T0 (BMJD(TDB))                    & 48689.14228(16) \\
  Distance                          & $48\pm5$pc \\
  $K_\mathrm{WD}$                    & $139.0\pm1.0$\kms \\
  $\gamma_\mathrm{WD}$               & $43.2\pm0.8$\kms \\
  $K_\mathrm{sec}$                   & $275.8\pm2.0$\kms \\
  $\gamma_\mathrm{sec}$               & $-4.4_{-0.9}^{+0.5}$\kms \\
  Mass ratio                        & $0.489\pm0.005$    \\
  Orbital separation                & $1.253\pm0.007$\RSUN \\
  Orbital inclination               & $77.7^\circ\pm0.5^\circ$ \\
  White dwarf mass                  & $0.782\pm0.013$\MSUN \\
  White dwarf radius                & $0.01068\pm0.00007$\RSUN \\
  White dwarf $\log{g}$             & $8.274\pm0.009$\\
  White dwarf effective temperature & 14,220$\pm350$K \\
  M star mass                      & $0.382\pm0.006$\MSUN \\
  M star radius sub-stellar        & $0.466\pm0.003$\RSUN \\
  M star radius polar              & $0.365\pm0.003$\RSUN \\
  M star radius backside           & $0.416\pm0.003$\RSUN \\
  M star radius volume-averaged    & $0.381\pm0.003$\RSUN \\
  \hline
\end{tabular}
\end{table}

The best fit models to the white dwarf eclipse in the $u'g'r'$ and $i'$ bands
are shown in Figure~\ref{fig:lc_fit}, along with their residuals. The results
from all four bands were consistent and the final stellar and binary
parameters are detailed in Table~\ref{tab:params}. The results are consistent
with those of \citet{odonoghue03}, but a factor of 4 more precise in
mass and almost 10 in radius. In fact, the radius measurement of the white
dwarf in QS\,Vir is the most precise, direct measurement of a white dwarf
radius ever. It is remarkable that we have measured the size of an 
object at nearly 50pc away to within 50km, and highlights the potential for
these kinds of systems to strongly test theoretical mass-radius relationships.
The measured masses of the two stars are also consistent with the results from
Roche tomography ($M_\mathrm{WD}=0.76\pm0.01$\MSUN,
$M_\mathrm{dM}=0.36\pm0.02$\MSUN).

An additional constraint on the stellar parameters comes from the gravitational
redshift of the white dwarf. For the measured parameters listed in
Table~\ref{tab:params} we would expect to measure a redshift for the white
dwarf of $45.8\pm0.6${\kms} (taking into account the redshift of the M star,
the difference in transverse Doppler shifts and the potential at the M star
owing to the white dwarf). Taking the difference between the systemic
velocities of the two stars from our UVES data gives a measured gravitational
redshift of $47.6\pm1.2${\kms}, which is fully consistent with our results. 

\section{Discussion} 

\subsection{The mass-radius relationship for white dwarfs} \label{sec:wdmr}

\begin{figure}
  \begin{center}
    \includegraphics[width=\columnwidth]{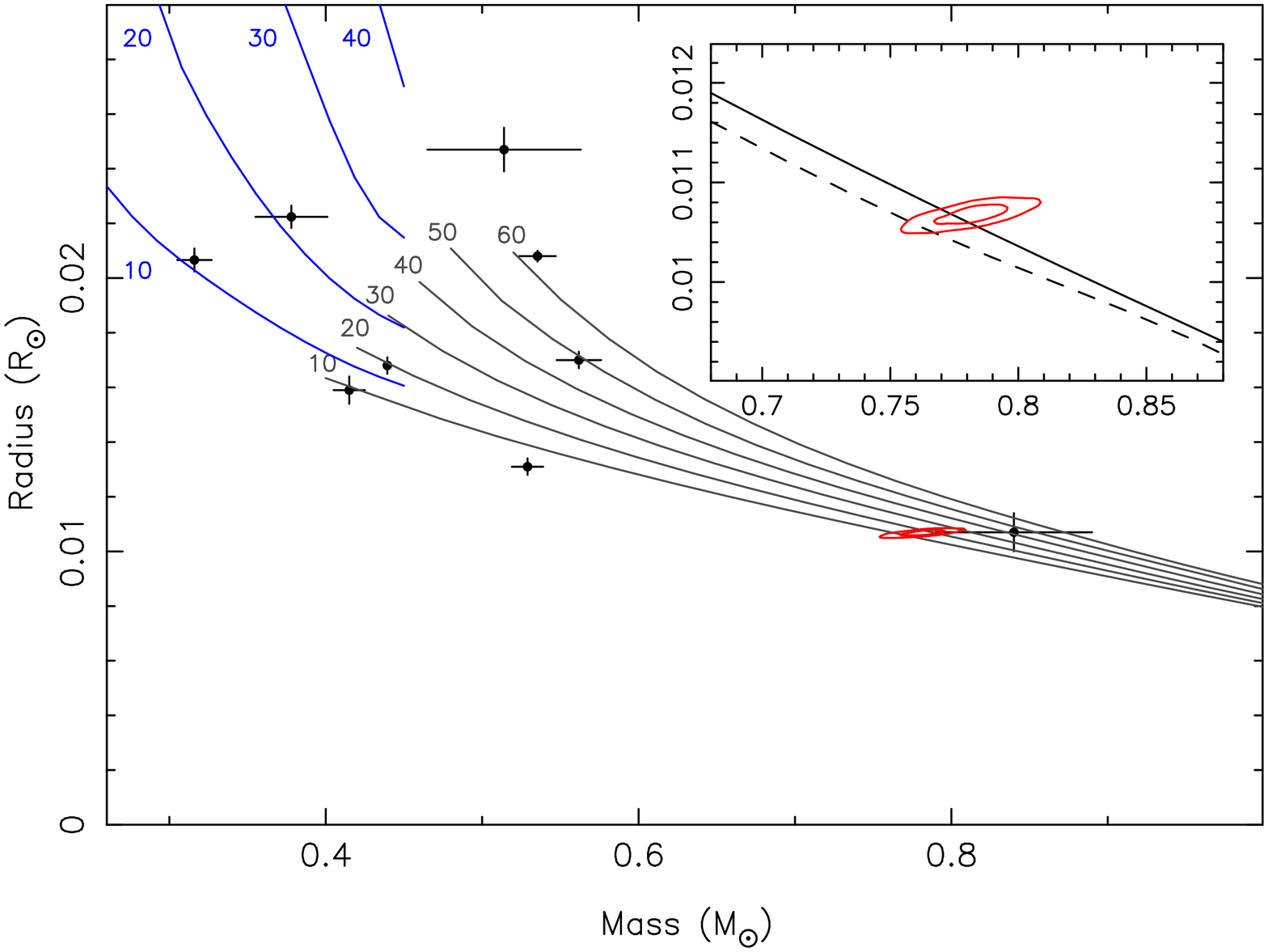}
    \caption{White dwarf mass-radius relation. We show the measured values for
      the white dwarf in QS\,Vir as red contours (68 and 95 percentile
      regions). Also shown as black points are other direct mass-radius
      measurements from eclipsing PCEB systems
      \citep{obrien01,pyrzas12,parsons10nn,parsons12hipr,parsons12cool,parsons12lowm,bours14,bours15}. Theoretical
    mass-radius relationships for carbon-oxygen core white dwarfs are shown in
  grey \citep{benvenuto99} and helium core white dwarfs in blue
  \citep{panei07}, labelled by temperature in thousands of Kelvin and for
  hydrogen envelope thicknesses of
  M$_\mathrm{H}$/M$_\mathrm{WD}=10^{-4}$. Inset we show a zoom of the measured
QS\,Vir parameters with a theoretical mass-radius relationship for a 14,200K
white dwarf with a thick hydrogen envelope
(M$_\mathrm{H}$/M$_\mathrm{WD}=10^{-4}$, solid line) and a thin envelope
(M$_\mathrm{H}$/M$_\mathrm{WD}=10^{-6}$, dashed line), showing that the
measured parameters are in excellent agreement with the theoretical models for
a white dwarf with a thick hydrogen envelope.}
  \label{fig:wdmr}
  \end{center}
\end{figure}

Figure~\ref{fig:wdmr} shows the mass-radius plot for white dwarfs and
highlights the location of the white dwarf in QS\,Vir. We also plot the
model-independent mass-radius measurements of several other white dwarfs, all
members of close, eclipsing binaries. Inset we show a zoom in on the white
dwarf in QS\,Vir compared to theoretical models. The measured values show
excellent agreement with a carbon-oxygen core white dwarf model of the same
temperature and with a thick hydrogen envelope
(M$_\mathrm{H}$/M$_\mathrm{WD}=10^{-4}$), although the uncertainty in the
measurements mean that thinner hydrogen layers cannot be ruled out. This is
consistent with the results from other white dwarfs in PCEBs, such as NN\,Ser
and GK\,Vir, which both possesses hydrogen envelope masses of
M$_\mathrm{H}$/M$_\mathrm{WD}=10^{-4}$ \citep{parsons10nn,parsons12hipr},
although several other eclipsing white dwarfs have thinner hydrogen envelopes
e.g. SDSS\,J0138-0016 (M$_\mathrm{H}$/M$_\mathrm{WD}=10^{-5}$,
\citealt{parsons12cool}) and SDSS\,J1212-0123
(M$_\mathrm{H}$/M$_\mathrm{WD}=10^{-6}$, \citealt{parsons12hipr}). Both white
dwarf components of the eclipsing binary CSS\,41177 appear to have thin
hydrogen layers as well (M$_\mathrm{H}$/M$_\mathrm{WD}<10^{-4}$,
\citealt{bours14}). The only other reliable white dwarf hydrogen envelope mass
measurement in a PCEB system is for the pulsating white dwarf in
SDSS\,J1136+0409, which was found to have an envelope mass of
M$_\mathrm{H}$/M$_\mathrm{WD}=10^{-4.9}$ \citep{hermes15}, implying that there
could be a substantial spread in the hydrogen envelope masses of white dwarfs
in PCEBs, possibly as a result of the common envelope phase itself.

The temperature and surface gravity of the white dwarf in QS\,Vir place it
close to the blue edge of the DA white dwarf instability strip. However, our
results are precise enough to firmly place it just outside the strip, which is
consistent with the null detection of pulsations in our extensive
high-precision photometry. 

\subsection{The mass-radius relationship for low-mass stars} \label{sec:mdmr}

\begin{figure}
  \begin{center}
    \includegraphics[width=\columnwidth]{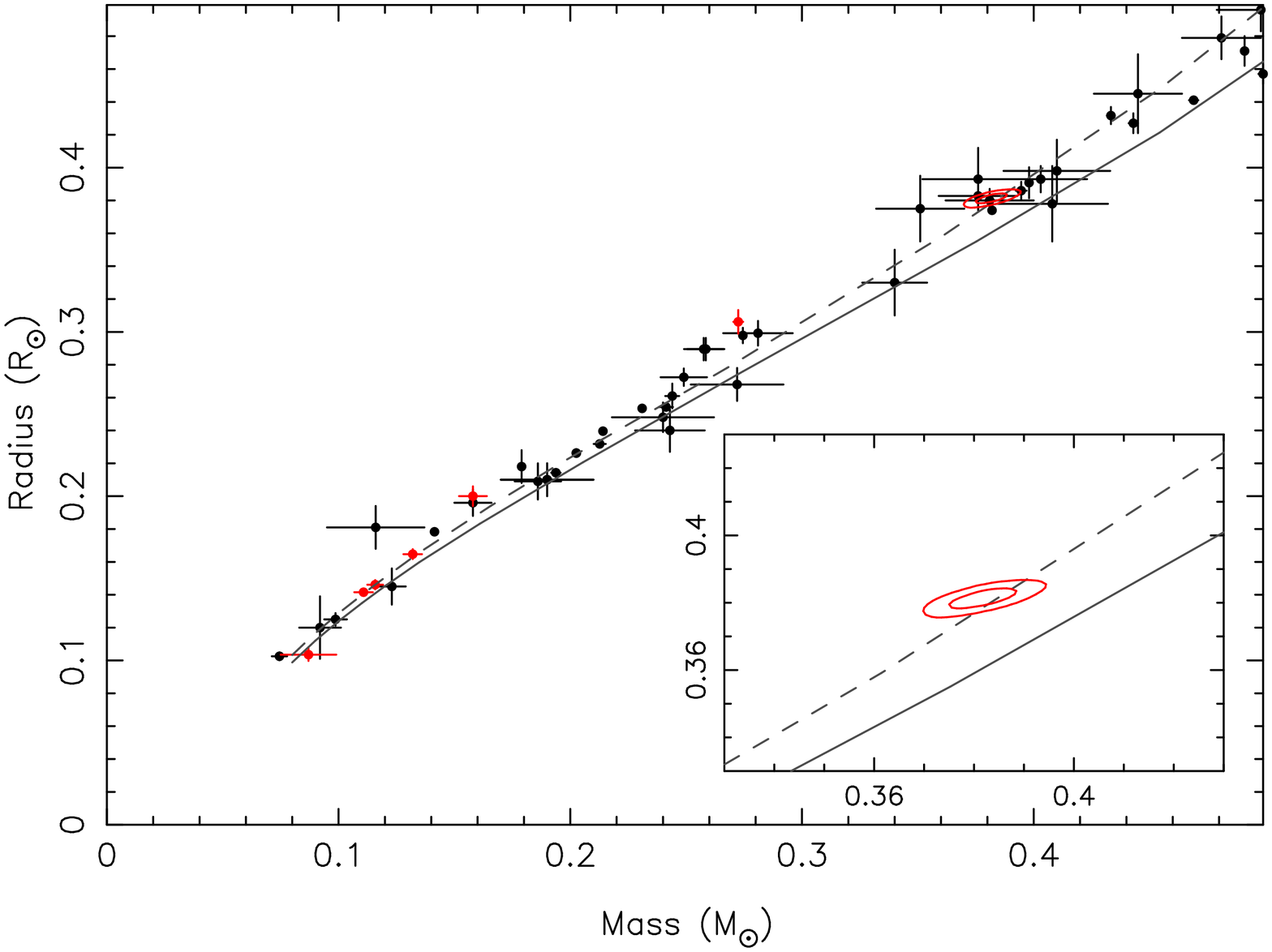}
    \caption{Low-mass star mass-radius relation. We show the measured values
      for the main-sequence star in QS\,Vir (volume-averaged radius) as red
      contours (68 and 95 percentile regions). Also shown as red points are
      other direct mass-radius measurements from eclipsing PCEB systems
      \citep{pyrzas12,parsons10nn,parsons12hipr,parsons12cool,parsons12lowm}.
      The black points are other direct mass-radius measurements taken from 
      \citet{gomezmc14}. The solid grey line is the theoretical mass-radius
      relation for a 1\,Gyr main-sequence star \citep{baraffe98}, while the
      grey dashed line is for an active star of the same age
      \citep{morales10}. Inset we show a zoom of the measured QS\,Vir
      parameters showing that the radius is consistent with a theoretical
      active star of the same mass.}
  \label{fig:mdmr}
  \end{center}
\end{figure}

In Figure~\ref{fig:mdmr} we show the mass-radius plot for low-mass
($<$0.5\MSUN) stars along with a number of directly measured objects. The
location of the M star in QS\,Vir is indicated by the contours, where we have
used the volume-averaged radius. Its mass and radius measurements are
consistent with the evolutionary models of \citet{morales10} for a 1\,Gyr
magnetically active star. Also highlighted in red are several other low-mass
star mass-radius measurements from PCEB systems. The M star in QS\,Vir is the
first low-mass star in a PCEB that is not fully convective
(M$_\mathrm{dM}>$0.35\MSUN) to have high-precision, model-independent mass and
radius measurements. While the majority of these measurements are consistent
with theoretical models, two systems (SDSS\,J1212-0123,
\citealt{parsons12hipr} and SDSS\,J1210+3347 \citealt{pyrzas12}) have measured
radii almost 10 per cent over-inflated compared to these same models. Any
possible explanation for why some of these low-mass stars are over-inflated,
while others aren't will require more high-precision mass-radius measurements
from PCEBs and more theoretical modelling of low-mass stars.

\subsection{The current evolutionary state of QS\,Vir} \label{sec:evo}

\begin{figure}
  \begin{center}
    \includegraphics[width=\columnwidth]{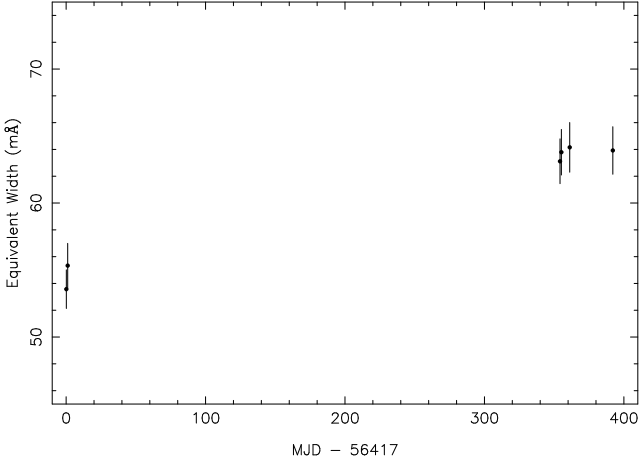}
    \caption{Equivalent width of the \Ion{Mg}{ii} 4481{\AA} absorption line
      from the white dwarf as a function of time. The strength of the line
      appears to be variable on year timescales, implying that the accretion
      rate of wind material on to the white dwarf is not constant.}
  \label{fig:mgii_ew}
  \end{center}
\end{figure}

The strong surface gravity of the white dwarf causes metals to sink out of the
photosphere on a short timescale. Therefore, the detection of the \Ion{Mg}{ii}
4481{\AA} absorption line implies that the white dwarf must be currently
accreting some material originating from the M star. Interestingly, the
strength of this absorption is variable as shown in
Figure~\ref{fig:mgii_ew}. It was slightly stronger in 2014 compared to
2013. This supports the idea that the accretion rate of material onto the
white dwarf is variable on year timescales and is unlikely to originate solely
from the wind of the M star, but is likely driven by flare and coronal
mass ejections (CME) events.

\citet{matranga12} detected clear X-ray eclipses, revealing
that the white dwarf dominated the X-ray flux rather than the M star,
supporting the idea of ongoing accretion. They determined an accretion
rate of $\dot{M} = 1.7 \times 10^{-13}\mathrm{M}_{\odot}\mathrm{yr}^{-1}$,
which would be one of the lowest accretion rates for a non-magnetic CV system
if this were the case, but would be particularly large for a detached system. 
The authors speculate that a stellar wind could be the source of the observed
mass transfer, invoking a magnetic `syphon' model, as found by \citet{cohen12}. 
\citet{matranga12} speculated that to maintain a more constant mass supply,
the wind could be supplemented by upward chromospheric flows, analogous to
spicules and mottles on the Sun.

Conversely, \citet{drake14} used the HST spectra taken by \citet{odonoghue03}
in 1999 to calculate an accretion rate of
$10^{-16}\mathrm{M}_{\odot}\mathrm{yr}^{-1}$, around one thousand times lower
than that found 6.5 years later by \citet{matranga12}. This accretion rate is
more consistent with a wind accretion, and is an order of magnitude lower than
the wind accretion rate inferred for similar pre-CVs
\citep{debes06,tappert11,parsons12hipr}. This supports the idea that CME and
prominence activity may provide a stochastic source of accretion material.

\begin{figure}
  \begin{center}
    \includegraphics[width=\columnwidth]{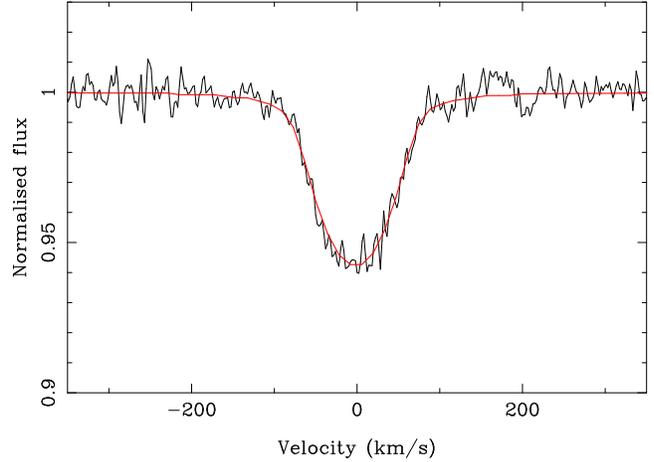}
    \caption{The \Ion{Mg}{ii} 4481{\AA} absorption line
      from the white dwarf made by combining all observations. Overplotted in
      red is a model with $V_\mathrm{rot}\sin{i}=69$\kms.}
  \label{fig:magab}
  \end{center}
\end{figure}

In \citet{parsons11} we used the \Ion{Mg}{ii} 4481{\AA} absorption line to rule
out the $V_\mathrm{rot}\sin{i}=400${\kms} rapid rotation of the white
dwarf determined by \citet{odonoghue03}. The long exposure time of that data
(1800\,s) meant that the line suffered from substantial smearing and
therefore it was consistent with having no measurable broadening and we
concluded that the white dwarf was not rapidly rotating. Our new data have
much shorter exposure times (180\,s) and can place better constraints on its
rotational velocity. In Figure~\ref{fig:magab} we show the \Ion{Mg}{ii}
4481{\AA} absorption line made by combining all the data, overplotted with the
best fit TLUSTY model \citep{hubeny95} with
$V_\mathrm{rot}\sin{i}=69$\kms. The white dwarf moves by a maximum of
$\sim$15{\kms} during each exposure, meaning that there is clearly some
additional broadening of the line. If this is due to the rotation of the white
dwarf then it implies a rotational period of only 700 seconds, extremely fast
for a detached system and implies that there may well have been a previous
stage of high mass-transfer. To spin up the white dwarf this much would
require an average accretion rate of
$\sim$$10^{-11}\mathrm{M}_{\odot}\mathrm{yr}^{-1}$ over the entire cooling age
of the white dwarf, meaning that QS\,Vir would almost certainly have to
have been a CV in the past. Alternatively, the white dwarf may posses a weak
magnetic field. A field strength of $\sim$$10^5$G is sufficient to Zeeman
split the \Ion{Mg}{ii} line by 69\kms, mimicking a broadening of the
line. This could also help explain the stability of the prominence features,
as the interaction of the fields of the two stars may create stable
regions. 

There are several pieces of evidence that could suggest that QS\,Vir is a
currently detached CV (either hibernating or genuinely detached, similar to a
CV crossing the gap). The fact that the binary is very close to Roche-lobe
filling, the white dwarf is relatively massive (0.782\MSUN, typical for white
dwarfs in CVs but relatively rare among PCEBs and pre-CVs
\citealt{zorotovic11}) and could be rapidly rotating, may indicate a previous
episode of high mass transfer. However, our results argue against this
interpretation. The RLFF of 96 per cent (97 per cent from the Roche tomography
analysis), means that the M star is still within its Roche lobe, and the fact
that its mass and radius are consistent with evolutionary models implies that
it is not inflated. In CVs above the orbital period gap the donor stars are
oversized by roughly 30 per cent \citep{knigge11} and would take at least
$10^7$\,yr (the thermal timescale of the M star) to relax back to their
equilibrium radius once becoming detached (i.e. if QS\,Vir is a detached CV,
then it must have been so for at least $10^7$\,yr). In this time the period
has decreased by $\sim$15 minutes (assuming classical magnetic braking,
\citealt{rappaport83}). Hence, if the system is in a hibernating state, then
the nova explosion that caused the system to detach would have had to increase
the period by $\sim$15 minutes, an unrealistic prospect. Its current period of
3.6\,h is also well above the standard period gap. Therefore, we conclude that
the system is more likely to be a pre-CV just about to fill its Roche-lobe,
and the white dwarf is weakly magnetic, mimicking the effects of rapid
rotation. However, confirmation of this will require spectropolarimetry around
the \Ion{Mg}{ii} 4481{\AA} line. 

Assuming classical magnetic braking and following the procedure outlined in
\citet{schreiber03}, we find that QS\,Vir will fill its Roche lobe at a period
of 3.363 hours, in 1.2\,Myr. However, the strength of magnetic braking is
highly uncertain. For example, replacing the formula given by Rappaport with
the normalised prescription used by \citet{davis08} results in a significantly
longer time until it fills its Roche lobe, i.e. 7.2\,Myr. Given that the
current cooling age of the white dwarf is 350\,Myr the probability of finding
a system as close to Roche-lobe filling as QS\,Vir is approximately half a per
cent, which is not implausible given that we now have more than 100 PCEBs with
measured periods.

\section{Conclusions}

We have combined high-resolution phase-resolved UVES spectroscopy and
high-speed ULTRACAM photometry of the eclipsing PCEB QS\,Vir to precisely
determine the binary and stellar parameters. We find that both the white dwarf
and its low-mass M star companion have radii fully consistent with
evolutionary models. The spectroscopy revealed large prominence structures
originating from the M star passing in front of both stars. Despite being
located in unstable locations within the binary, these prominences appear to
be long-lived, lasting more than a year. Using Roche tomography we determined
that the M star is covered with a large number of starspots, preferentially
located on its inner hemisphere, facing the white dwarf. They too appear to be
long-lived and may well be related to the prominences (e.g. located close to
where the prominences are anchored to the M star's surface).

The M star in QS\,Vir fills 96 per cent of its Roche lobe and the system
appears to be a pre-cataclysmic binary, yet to have initiated mass transfer
via the inner Lagrange point, rather than a hibernating system. However, the
white dwarf may be rapidly rotating, or weakly magnetic, meaning that its past
evolution is still uncertain. It will become a cataclysmic variable system in
1.2\,Myr with a period above the period gap. 

\section*{Acknowledgments}

We thank the anonymous referee for helpful comments and suggestions.
The authors would like to dedicate this paper to the memory of Darragh
O'Donoghue, who will be missed as a brilliant and inspiring colleague
and who wrote the excellent discovery paper for QS\,Vir. 
SGP and MZ acknowledge financial support from FONDECYT in the form of grant
numbers 3140585 and 3130559. CAH acknowledges the Queen's University Belfast
Department of Education and Learning PhD scholarship. ULTRACAM, TRM, CAW, VSD
and SPL are supported by the Science and Technology Facilities Council (STFC),
TRM and DS acknowledge grant number ST/L000733. The research leading to these
results has received funding from the European Research Council under the
European Union's Seventh Framework Programme (FP/2007-2013) / ERC Grant
Agreement n. 320964 (WDTracer). MRS thanks for support from FONDECYT (1141269)
and Millennium Science Initiative, Chilean ministry of Economy: Nucleus
P10-022-F. The results presented in this paper are based on observations
collected at the European Southern Observatory under programme IDs 090.D-0277
and 093.D-0096.

\bibliographystyle{mn_new}
\bibliography{qsvir}

\label{lastpage}

\end{document}